\documentclass[journal]{IEEEtran}

\ifCLASSINFOpdf
\else
\fi

\usepackage{graphicx}
\usepackage{subfigure}

\usepackage{graphicx}
\usepackage{url}
\usepackage{bm}
\usepackage[comma,numbers,square,sort&compress]{natbib}
\usepackage{hyperref} 
\usepackage{amsmath,amssymb,amsthm}  
\usepackage{paralist}

\usepackage{algorithm}
\usepackage{algorithmic}      
\usepackage{multirow}
\renewcommand{\algorithmicrequire}{\textbf{Initialization:}}
\renewcommand{\algorithmicensure}{\textbf{Iteration:}}
\renewcommand{\algorithmiclastcon}{\textbf{Output:}} 

\usepackage{cleveref}
\Crefname{figure}{Fig.}{Figs.}

\crefname{assume}{assumption}{assumptions}

\newtheorem{lemma}{Lemma}

\newtheorem{theorem}{Theorem}

\newtheorem{assume}{Assumption}

\newcommand{\e}{\begin{equation}}
\newcommand{\ee}{\end{equation}}
\newcommand{\en}{\begin{equation*}}
\newcommand{\een}{\end{equation*}}
\newcommand{\eqn}{\begin{eqnarray}}
\newcommand{\eeqn}{\end{eqnarray}}
\newcommand{\bmat}{\begin{bmatrix}}
\newcommand{\emat}{\end{bmatrix}}
\newcommand{\BIT}{\begin{itemize}}
\newcommand{\EIT}{\end{itemize}}

\newcommand{\Trans}{{\mathrm{T}}}  %
\newcommand{\HTrans}{{\mathrm{H}}}  %

\newcommand{\defequ}{\triangleq}

\newcommand{\xmath}[1]{\ensuremath{#1}\xspace}
\newcommand{\blmath}[1]{\bm{\mathrm{#1}}}

\newcommand{\uvg}{\blmath{g}}

\newcommand{\uvm}{\blmath{m}}

\newcommand{\uvs}{\blmath{s}}

\newcommand{\uvu}{\blmath{u}}
\newcommand{\uvv}{\blmath{v}}
\newcommand{\uvw}{\blmath{w}}
\newcommand{\uvx}{\blmath{x}}
\newcommand{\uvy}{\blmath{y}}
\newcommand{\uvz}{\blmath{z}}

\newcommand{\umA}{\blmath{A}}
\newcommand{\umB}{\blmath{B}}

\newcommand{\umD}{\blmath{D}}

\newcommand{\umF}{\blmath{F}}

\newcommand{\umH}{\blmath{H}}
\newcommand{\umI}{\blmath{I}}

\newcommand{\umP}{\blmath{P}}

\newcommand{\umS}{\blmath{S}}

\newcommand{\umW}{\blmath{W}}

\newcommand{\vtheta}{\bm \theta}

\newcommand{\proxW}[3]{{\mathrm{prox}}_{#1}^{#2}{#3}}

\newcounter{oursection}

\usepackage{xspace}
\newcommand{\Dsig}{\xmath{\umD}}

\usepackage{color}

\usepackage{silence}
\usepackage{soul}
\theoremstyle{plain} 

\newtheorem{thm}{Theorem}[section] 
\newcommand{\thistheoremname}{}
\newtheorem{genericthm}[thm]{\thistheoremname}

\newtheorem*{genericthm*}{\thistheoremname}
\newenvironment{namedthm*}[1]
  {\renewcommand{\thistheoremname}{#1}%
   \begin{genericthm*}}
  {\end{genericthm*}}

\usepackage{mathtools, cuted}
\usepackage{relsize}
\usepackage[export]{adjustbox}
\usepackage[dvipsnames]{xcolor}
\usepackage[skins]{tcolorbox}

\definecolor{darkred}{rgb}{0.6,0,0}
\definecolor{darkgreen}{rgb}{0,0.5,0}
\definecolor{darkblue}{rgb}{0,0,0.5}
\definecolor{goldenrod}{rgb}{0.85, 0.65, 0.13}
\definecolor{goldenbrown}{rgb}{0.6, 0.4, 0.08}

\definecolor{convindigo}{RGB}{178,200,255}
\definecolor{actcoral}{RGB}{255,160,160}
\definecolor{poolamber}{RGB}{255,220,150}
\definecolor{resteal}{RGB}{160,225,200}

\usepackage{tikz,pgfplots}
\pgfplotsset{compat=1.5.1}
\usepackage{adjustbox}
\usetikzlibrary{intersections, pgfplots.fillbetween, positioning, shapes.multipart, calc, fit, backgrounds}
\usepgfplotslibrary{groupplots}
\newcommand{\stepsize}{\alpha}
\usepackage{upgreek}

\usepackage{multicol}

\newcommand{\MRcb}[1]{{\color{black}#1}}

\newcommand{\MZMSCHR}{MMESHR$1$\xspace}

\long\def\red#1{\bgroup\color{red}#1\egroup}

\begin{document}
\title{Convergent Complex Quasi-Newton Proximal Methods
for Gradient-Driven Denoisers
\\
in Compressed Sensing MRI Reconstruction}

\author{ 
Tao Hong \IEEEmembership{Member, IEEE}, Zhaoyi Xu, Se Young Chun \IEEEmembership{Member, IEEE}, Luis Hernandez-Garcia,
\\ and
Jeffrey A. Fessler \IEEEmembership{Fellow, IEEE}



\thanks{T. Hong is with the Oden Institute for Computational Engineering and Sciences,
University of Texas at Austin, Austin, TX 78712, USA (Email: \texttt{tao.hong@austin.utexas.edu}). TH was with the Department of Radiology, University of Michigan, Ann Arbor, MI 48109, USA. TH was partially supported by NIH grant R01NS112233.
}
\thanks{Z. Xu is with the Department of Mechanical Engineering,
University of Michigan, Ann Arbor, MI 48109,  USA
(Email: \texttt{zhaoyix@umich.edu}).}

\thanks{S. Chun is with the Department of ECE, INMC \& IPAI, SNU, Seoul, 08826, South Korea (Email: sychun@snu.ac.kr). SC was partially supported by IITP grant [No.RS-2021-II211343, Artificial Intelligence Graduate School Program (Seoul National University)] and NRF grant (No. RS-2025-02263628), all funded by the Korea government(MSIT).
}
\thanks{L. Hernandez-Garcia is with the Department of Radiology, University of Michigan, Ann Arbor, MI 48109, USA
(Email: \texttt{hernan@umich.edu}). LH was partially supported by NIH grant R01NS112233.}

\thanks{J. Fessler
is with the EECS Department,
University of Michigan, Ann Arbor, MI 48109, USA 
(Email: \texttt{fessler@umich.edu}).
Supported in part by NIH grants
R01EB035618 and R21EB034344.
}
}


%
%

\markboth{}
{Shell \MakeLowercase{\mrmit{et al.}}: Bare Demo of IEEEtran.cls for IEEE Journals}
%


\maketitle

\begin{abstract}
In compressed sensing (CS) MRI, model-based methods are pivotal to achieving accurate reconstruction. One of the main challenges in model-based methods is finding an effective prior to describe the statistical distribution of the target image.
Plug-and-Play (PnP) and REgularization by Denoising (RED) are two general frameworks
that use denoisers as the prior.
While PnP/RED methods with convolutional neural network (CNN) based denoisers outperform classical hand-crafted priors in CS MRI, their convergence theory relies on assumptions that do not hold for practical CNN models.
The recently developed gradient-driven denoisers offer a framework that bridges the gap between practical performance and theoretical guarantees. However, the numerical solvers for the associated minimization problem remain slow for CS MRI reconstruction.
This paper proposes a complex quasi-Newton proximal method
that achieves faster convergence than existing approaches.
To address the complex domain in CS MRI,
we propose a modified Hessian estimation method
that guarantees Hermitian positive definiteness.
Furthermore, we provide a rigorous convergence analysis of the proposed method
for nonconvex settings.
Numerical experiments on both Cartesian and non-Cartesian sampling trajectories
demonstrate the effectiveness and efficiency of our approach.
\end{abstract}

\begin{IEEEkeywords}
CS MRI, gradient-driven denoiser, second-order, convergence, complex domain,
spiral and radial acquisitions.
\end{IEEEkeywords}

%


\IEEEpeerreviewmaketitle

\section{Introduction}
\label{sec:Introduction}
\IEEEPARstart{M}{AGNETIC} resonance imaging (MRI)
is a non-invasive imaging technique that generates images
of the internal structures of the body \cite{liang2000principles}.
MRI is widely used in clinical settings for disease diagnosis,
treatment guidance, and functional and advanced imaging,
among other applications \cite{brown2011mri}.
In practice, MRI scanners acquire k-space data,
which are the Fourier components of the desired images.
The acquisition procedure is slow, leading to patient discomfort,
increased motion artifacts, reduced clinical efficiency, and other issues.
To accelerate the acquisition, modern MRI scanners use multiple coils (parallel imaging)
to acquire less Fourier components.
The  parallel imaging  technique incorporates additional spatial information
that can help significantly reduce MRI acquisition time
\cite{pruessmann1999sense,griswold2002generalized}.
Moreover, by combining with compressed sensing (CS)~\cite{lustig2007sparse},
one can acquire even fewer Fourier components,
further accelerating the acquisition process.
However, the reconstruction in CS MRI requires iterative solvers
for the following composite minimization problem:
\begin{equation}
	\label{eq:CSMRIReco:origP}
\uvx^*=\arg\min_{\uvx\in\mathbb C^N} F(\uvx) \equiv \underbrace{\frac{1}{2}\|\umA\uvx-\uvy\|_2^2}_{h(\uvx)}+\lambda\, f(\uvx),
\end{equation}
where $\umA\in\mathbb C^{MC\times N}$ denotes the forward model
specifying the mapping from the image $\uvx\in \mathbb C^N$
to the k-space data  $\uvy\in\mathbb C^{ML}$,
and $f(\uvx)$ refers to the regularizer
describing the prior information about $\uvx$.
In practice, we have $M\ll N$ due to downsampling.
\MRcb{The trade-off parameter $\lambda > 0$ balances the data-fidelity term $h(\uvx)$ and the regularizer $f(\uvx)$.
Here $C > 1$ denotes the number of coils.
The system matrix $\umA$ is a stack of submatrices $\umA_c$ such that $\umA=\left[\umA_1;\umA_2;\cdots;\umA_C\right]$.} The submatrices $\umA_c$ are defined as $\umA_c\in\mathbb C^{M\times N}=\umP\umF\umS_c$ where $\umP$ is the downsampling pattern, $\umF$ defines the (non-uniform) Fourier transform,
and $\umS_c$ denotes the coil sensitivity map for $c$th coil, which is patient specific.   

The data fidelity term $h(\uvx)$ enhances the data consistency. Since the k-space data is highly downsampled, the regularizer $f(\uvx)$ is required to stabilize the solution. The choice of $f$ can significantly affect the reconstruction quality. Classical hand-crafted regularizers  have proven effective for MRI reconstruction including  wavelets~\cite{guerquin2011fast,zibetti2018monotone}, total variation (TV)~\cite{rudin1992nonlinear}, a combination of wavelet and TV \cite{lustig2007sparse,hong2024complex}, dictionary learning~\cite{aharon2006k,ravishankar2011mr}, and low-rank methods~\cite{dong2014compressive}, to name a few.
See \cite{fessler:10:mbi, fessler2020optimization}
for a review of various choices for $f$. 

Over the past decade, deep learning (DL) has attracted significant attention for MRI reconstruction because of its superior performance \cite{heckel2024deep}. Unlike hand-crafted regularizers, DL learns complex image priors directly from large amounts of data. The promising DL-based methods for MRI reconstruction include end-to-end learning frameworks~\cite{Wang2016.etal} and physics-informed deep unrolling methods~\MRcb{\cite{aggarwal2018modl,gilton2021deep,ramzi2022nc,wang2023one,shi2023provable}}. More recently, generative models have emerged as powerful tools for learning priors in MRI reconstruction, gaining substantial interest~\cite{song2021solving,chung2022score}.  

An alternative framework to DL is Plug-and-Play (PnP)/REgularization by Denoising (RED)~\cite{venkatakrishnan2013plug,romano2017little}, which leverages the most effective denoisers, such as BM3D~\cite{dabov2007image} or DnCNN~\cite{zhang2017beyond}, achieving outstanding performance in various imaging tasks~\MRcb{\cite{sreehari2016plug, Ono2017, Meinhardt.etal2017, Buzzard.etal2018, shi2020deep,zhang2021plug}}. Compared to the end-to-end and unrolling DL approaches, which are typically designed for a predefined imaging task and rely on training with massive amounts of data, PnP/RED can be easily adapted to specific applications without requiring retraining. This capability is especially advantageous for addressing CS MRI problems, where sampling patterns, coil sensitivity maps, and image resolutions can vary greatly between scans.  Detailed discussions about using PnP for MRI reconstruction are found in~\cite{ahmad2020plug}.
The following subsections first introduce background on PnP/RED priors
and related theoretical work.
We then discuss the gradient-driven denoisers framework
and the associated minimization problem.

\subsection{Inverse Problems with PnP/RED Priors}

Proximal algorithms~\cite{parikh2014proximal} are a class of iterative methods
for solving \eqref{eq:CSMRIReco:origP}.
At $k$th iteration, the proximal gradient method (PGM) is expressed as
\begin{equation}
\label{eq:ISTA}    
\uvx_{k+1}  =  \proxW{\stepsize \lambda f}{\umI}{(\uvx_k-\stepsize \nabla h(\uvx_k))}, 
\end{equation}
where $\stepsize\in\mathbb R,~\alpha > 0 $ is the stepsize,   
$\nabla h (\cdot)$ denotes the gradient of $h(\cdot)$, and
$\proxW{\stepsize \lambda f}{\umW}{(\cdot)}$
denotes the weighted proximal mapping (WPM)
defined as
\begin{equation}
    \label{eq:prox}
     \proxW{\stepsize \lambda f}{\umW}{(\cdot)}\defequ \arg\min_{\uvx\in\mathbb C^N}\frac{1}{2}\|\uvx-\cdot \|_{\umW}^2+ \stepsize \lambda f(\uvx),
\end{equation}
where $\umW\in\mathbb C^{N\times N},~\umW\succ0$ is a Hermitian positive definite matrix and $\|\uvx\|_{\umW}^2=\uvx^\HTrans \umW\uvx$. Here, $\HTrans$ denotes the Hermitian transpose operator.
Clearly, $\proxW{\stepsize \lambda f}{\umI}{(\cdot)}$ represents the classical proximal operator \cite{parikh2014proximal}.
Since $\proxW{\stepsize \lambda f}{\umI}{(\cdot)}$ can be interpreted as a denoiser,
the PnP-ISTA algorithm
simply replaces the proximal operator with an arbitrary denoisier $\Dsig(\cdot)$. In practice,
the sequence $\{\uvx_k\}$
generated by PnP-ISTA generally converges slowly.
To address this issue, Pendu et al.~\cite{pendu2023preconditioned} introduced a preconditioned PnP-ADMM algorithm
that uses a diagonal matrix as the preconditioner.
More recently, Hong et al.~\cite{hong2024provable}
presented a provably convergent
preconditioned PnP framework
that supports general preconditioners 
and demonstrates fast convergence in CS MRI reconstruction.


An alternative to PnP, RED~\cite{romano2017little,hong2019acceleration,reehorst2018regularization}
introduces an explicit regularization term based on a denoiser, i.e.,
$f(\uvx)=\frac{1}{2}\uvx^\Trans \left[\uvx-\Dsig(\uvx)\right]$.
Here, $\Trans$ denotes the transpose operator.
Romano et al. \cite{romano2017little} demonstrated that
if the denoiser satisfies the local homogeneity property,
the gradient of $f(\uvx)$ in RED can be expressed as $\uvx-\Dsig(\uvx)$.
Since $f(\uvx)$ in RED is differentiable, various iterative methods,
such as gradient descent, proximal methods, and quasi-Newton methods,
can be applied to solve RED,
provided that $h(\uvx)$ is also differentiable.
To accelerate convergence in RED, several techniques have been adopted,
such as vector extrapolation \cite{hong2019acceleration},
fast proximal methods \cite{reehorst2018regularization},
and weighted proximal methods \cite{hong2020solving}, among others.

While PnP/RED has demonstrated significant empirical success,
theoretical research on its convergence continues to be an active area of study, see~\MRcb{\cite{chan2017plug,romano2017little,reehorst2018regularization,Buzzard.etal2018,Sun.etal2019a,ryu2019plug,xu2020provable,shi2025provably,gavaskar2021plug,sun2021scalable, liu2021recovery, cohen2021regularization,kamilov2023plug}}.
These studies assume either that the denoiser approximates the MAP or MMSE estimator,
\MRcb{that $\|\Dsig(\uvx)-\uvx\|$ is upper bounded},
or that it is nonexpansive, satisfying
\begin{equation}
\label{eq:DenoiserLip}
	\|\Dsig(\uvx_1)-\Dsig(\uvx_2)\|\leq  \|\uvx_1-\uvx_2\|.
\end{equation}
However, it is known that if the Jacobian of such estimators is well-defined, it must be symmetric \cite{reehorst2018regularization,gribonval2020characterization,gribonval2021bayesian}. However, RED/PnP frequently achieves state-of-the-art performance with denoisers that do not satisfy this condition,
such as Convolutional Neural Network (CNN) based denoisers.
So RED/PnP cannot be explained as an optimization-based solver.
Although optimization-free explanations for RED and PnP exist,
understanding their behaviors and characterizing their solutions remains challenging.
Moreover, most commonly used denoisers do not satisfy~\eqref{eq:DenoiserLip}.
Another approach \MRcb{\cite{ryu2019plug,terris2020building,pramanik2023memory}}
is to train a denoiser
by incorporating a regularization term in training to constrain its Lipschitz constant.
However, this approach remains an open challenge,
as there is currently no practical way to strictly guarantee
that the trained denoiser is nonexpansive \MRcb{\cite{aziznejad2020deep,bredies2024learning}}.

\subsection{Inverse Problems with gradient-driven Denoisers}

To address the aforementioned challenges and bridge the gap
between theoretical assumptions and practical performance in PnP/RED,
recent works~\cite{cohen2021has, hurault2021gradient, fang2023s, chaudhari2024gradient}
have proposed constructing gradient-driven denoisers.
In this framework, the denoiser is given
by the difference between the noisy image $\uvx$
and the gradient of a scalar-valued function $f_{\bm\uptheta}(\uvx)$, i.e.,
\begin{equation}
\label{eq:grad_denoiser}
    \umD_{\bm\uptheta}(\uvx) \equiv \uvx - \nabla_{\uvx} f_{\bm\uptheta}(\uvx).
\end{equation}
In practice, $f_{\bm\uptheta}(\uvx)$ is constructed with a scalar-output CNN
and trained so that
$\uvx - \nabla_{\uvx} f_{\bm\uptheta}(\uvx)$ serves as a denoiser.
Here ${\bm\uptheta}$ denotes the trained network parameters.
In medical imaging reconstruction, interpretability, reliability,
and provable numerical algorithms are crucial,
as the reconstructed images are commonly used for disease diagnosis.
Therefore, it is essential to understand how the underlying algorithms function.
The use of gradient-driven denoisers for CS MRI reconstruction
enables the integration of deep learning techniques
while maintaining interpretability and reliability.
Specifically, we aim to solve the following optimization problem
to recover the latent image:
\begin{equation}
\label{eq:CSMRIReco:origP:GradDenoiser}
{\uvx}^* = \arg\min_{\uvx \in \mathcal{C}}
F(\uvx) \equiv \underbrace{\frac{1}{2} \|\umA \uvx - \uvy\|_2^2}_{h(\uvx)}
+ \lambda f_{\bm\uptheta}(\uvx),
\end{equation}
where $\mathcal C$ is a closed convex set of $\mathbb C^N$.
For notational simplicity, we ignore $\bm\uptheta$ in $f_{\bm\uptheta}(\uvx)$
and simply write $f(\uvx)$ hereafter.
Similarly, we use $\nabla f(\uvx)$ to denote $\nabla_{\uvx} f(\uvx)$.
Once the trade-off parameter $\lambda$ is chosen,
we fix it throughout the minimization.
Therefore, we absorb $\lambda$ into $f$ hereafter
and write $f(\uvx)$ instead of $\lambda f(\uvx)$.

Although both $f(\uvx)$ and $h(\uvx)$ are differentiable,
$f(\uvx)$ may be nonconvex.
Therefore, the underlying numerical algorithms
for solving \eqref{eq:CSMRIReco:origP:GradDenoiser}
should be theoretically sound
and applicable to nonconvex settings.
Cohen et al. \cite{cohen2021has} employed the projected gradient method
with a line search strategy to solve \eqref{eq:CSMRIReco:origP:GradDenoiser}
and established its convergence
under the assumption that the gradient of $f(\uvx)$ is Lipschitz continuous.
Hurault et al.~\cite{hurault2021gradient} applied the proximal gradient method,
where at the $k$th iteration, $\uvx_{k+1}$ is updated as follows:
\begin{equation}
\label{eq:ISTAUpdate:origP:GradDenoiser}
\uvx_{k+1} = \proxW{\stepsize_k h+\iota_{\mathcal C}}{\umI}{\big (\uvx_k-\stepsize_k \nabla f(\uvx_k)\big )},
\end{equation}
where $\stepsize_k$ represents the stepsize and $ \iota_{\mathcal C}(\uvx)$
denotes the characteristic function such that
\begin{equation}
\label{eq:characteristic}
\iota_{\mathcal C}(\uvx) =
\begin{cases}
    0, & \text{if } \uvx \in \mathcal{C}, \\
    +\infty, & \text{otherwise}.
\end{cases}
\end{equation}
Similar to \cite{cohen2021has},
the convergence is established under the same assumption
and $\stepsize_k$ is determined using a line search strategy. Note that both methods were tested on image deblurring and super-resolution tasks
in the \emph{real} domain.
A direct extension of these methods to CS MRI reconstruction
requires hundreds of iterations to converge,
limiting their practical applicability.
Therefore, improving the convergence speed is essential for CS MRI reconstruction with gradient-driven denoisers.

\MRcb{Alternatively, Tan et al.~\cite{tan2023provably}
adopted the method proposed in~\cite{stella2017forward} to solve the following problem,
instead of \eqref{eq:CSMRIReco:origP:GradDenoiser},
by exploiting second-order information:
\begin{equation}
	\label{eq:PnPLBFGSOptPro}
	\min_{\uvx \in \mathbb{R}^N} h(\uvx)
	+ \lambda \underbrace{\left( f(\umD^{-1}(\uvx)) 
	- \tfrac{1}{2}\|\umD^{-1}(\uvx) - \uvx\|_2^2 \right)}_{\bar f(\uvx)},
\end{equation}
where $\umD(\uvx) = \uvx - \nabla f(\uvx)$ denotes the denoiser.  
Their algorithmic routine,
dubbed PnP-LBFGS,
treats
$\umD(\uvx)$ as a proximal mapping of $\lambda\bar f(\uvx)$.  
However, PnP-LBFGS requires the Lipschitz constant of $\nabla f$ to be upper bounded by one,
which is challenging to guarantee in practice.  
Furthermore, Tan et al.~only considered image deblurring and super-resolution problems
for \emph{real-valued} images,
and it remains unclear how convex constraints could be incorporated into their framework.
}

\subsection{Contributions and Roadmap}
\label{sec:ConRoad}

Motivated by recent work
demonstrating the potential advantages of using second-order information
for acceleration in many applications 
\cite{schmidt2009optimizing,lee2014proximal,chouzenoux2014variable,bonettini2016variable,karimi2017imro,becker2019quasi,hong2020solving,repetti2021variable,hong2024complex,zhang2025proximal},
we propose a \emph{complex} quasi-Newton proximal method (CQNPM)
that incorporates additional Hessian information at each iteration
to accelerate the convergence of solving \eqref{eq:ISTAUpdate:origP:GradDenoiser}.
In contrast to \cite{hong2024complex},
we estimate the Hessian matrix of $ f(\uvx)$ rather than that of $h(\uvx)$. Although $f(\uvx)$ is differentiable,
it is nonconvex and $\uvx$ is \emph{complex}, which introduces new challenges
in estimating a Hermitian positive definite Hessian matrix.
To address these challenges,
we develop an approach that enforces the estimated Hessian matrix
to be Hermitian positive definite.

The main contributions of this paper are summarized as follows: 
\begin{itemize}
\item
We extend the gradient-driven denoisers
to complex-valued CS MRI reconstruction.
Moreover, we propose a \emph{complex} quasi-Newton proximal approach
to efficiently solve the associated minimization problem.
	
\item
We propose a modified Hessian estimation method
that enforces a Hermitian positive definite Hessian matrix.
Moreover, we provide a rigorous convergence analysis of the proposed approach
under the nonconvex settings.
\item
We extensively validate the performance of our method
using both Cartesian and non-Cartesian sampling trajectories on brain and knee images.
\end{itemize}

The rest of this paper is organized as follows.
\Cref{sec:proposedMethod} introduces our proposed method
and explains how to estimate a Hermitian positive definite Hessian matrix.
In addition,
\Cref{sec:convAnalysis} provides a rigorous convergence analysis.
Finally,
\Cref{sec:numericalExp} presents
numerical experiments that study the performance of our method
and validate the theoretical analysis. \MRcb{Supplementary material is also provided, including the validation of \Cref{assum:LipGrad}, a comparison with PnP-LBFGS, and additional experimental results.}


\section{Proposed Method}
\label{sec:proposedMethod}

This section first introduces our approach
for addressing~\eqref{eq:CSMRIReco:origP:GradDenoiser}.
We then describe how to estimate a Hermitian positive definite Hessian matrix. Finally, we present an efficient strategy for solving the associated WPM.


Given an estimated Hessian matrix $\umB_k$, at the $k$th iteration,
we update the next iterate by solving a WPM, i.e.,
\begin{equation}
\label{eq:MainUpdata}
\uvx_{k+1}=\proxW{\stepsize_k h+\iota_\mathcal C}{\umB_k}{\big(\uvx_k-\stepsize_k \,\umH_k\nabla f(\uvx_k)\big)},
\end{equation}
where $\umH_k=\umB_k^{-1}$.
Unlike the usual PGM in \eqref{eq:ISTA}
that uses the proximal mapping of the \emph{regularizer},
the WPM in \eqref{eq:MainUpdata}
uses the proximal mapping of the \emph{data term} $h$
plus the characteristic function in \eqref{eq:characteristic}. \Cref{alg:PropMethod} summarizes the proposed method.

To compute a Hermitian positive definite
\MRcb{approximation} to the Hessian matrix,
we introduce a modified memory-efficient self-scaling Hermitian rank-$1$ method (\MZMSCHR)
that extends the method proposed in 
\cite{osborne1999new,curtis2016self,wang2019stochastic,hong2024provable}. 
\Cref{alg:ModfiedZero:SR1} presents the detailed steps of \MZMSCHR. 
The operator $\Re(\cdot)$ in \Cref{alg:ModfiedZero:SR1}
denotes an operation that extracts the real part.
\MRcb{Without the operator $\Re(\cdot)$, \Cref{alg:ModfiedZero:SR1}
reduces to the Hessian approximation method proposed in~\cite{hong2024complex}.}
For the problems in~\cite{hong2024complex},
we estimated the Hessian matrix for $h(\uvx)$
so that $\langle \uvm_k,\uvs_k\rangle$ is guaranteed to be real for all iterations.
However, $\langle \uvm_k,\uvs_k\rangle$ becomes complex in our problem setting,
leading to a non-Hermitian Hessian approximation.
\MRcb{Let $\bar{\umB}_k$ denote the approximate Hessian matrix
obtained using the approach proposed in~\cite{hong2024complex}.
To obtain a Hermitian positive definite Hessian approximation,
we seek a $\umB_k$ that is the nearest Hermitian approximation to $\bar{\umB}_k$
in terms of Frobenius norm minimization,
resulting in
$\umB_k=(\bar{\umB}_k + \bar{\umB}_k^\HTrans)/2$.}
In practice, this step is equivalent to applying $\Re(\cdot)$ at steps
\ref{alg:ModfiedZero:SR1:stepv_k}, \ref{alg:ModfiedZero:SR1:steptau_k},
and \ref{alg:ModfiedZero:SR1:stepInnerrhok} in \Cref{alg:ModfiedZero:SR1}.
\Cref{lemma:boundedHessian} shows that
the \MRcb{approximate} Hessian matrices generated by \Cref{alg:ModfiedZero:SR1}
are always Hermitian positive definite,
even if $f$ is nonconvex.

\begin{algorithm}[t]        
\caption{Complex Quasi-Newton Proximal Method (CQNPM)}    
\label{alg:PropMethod} 
\begin{algorithmic}[1]
\REQUIRE $\uvx_1$ and stepsize $\stepsize_k > 0$
\ENSURE 
\FOR {$k=1,2,\dots$}
\STATE Set $\umH_k$ and $\umB_k$ using \Cref{alg:ModfiedZero:SR1}
\STATE $\uvx_{k+1} \leftarrow \proxW{\stepsize_k\,h+\iota_\mathcal C}
{\umB_k}{\big(\uvx_k-\stepsize_k\,\umH_k\nabla f(\uvx_k)\big)}$\label{alg:PropMethod:WPG}
\ENDFOR
\end{algorithmic}
\end{algorithm}


Computing the WPM in~\eqref{eq:MainUpdata} is equivalent
to solving the following minimization problem:
\begin{equation}
	\label{eq:MainUpdata:WPM:Open}
\arg \min_{\bar{\uvx}\in\mathcal C}
G_k(\bar{\uvx})\equiv \frac{1}{2}\left( \big\|\bar{\uvx}-\uvw_k\big\|_{\umB_k}^2+\stepsize_k\big\|\umA\bar{\uvx}-\uvy\big\|_2^2\right),
\end{equation}
where $\uvw_k = \uvx_k-\stepsize_k\umH_k\nabla f(\uvx_k)$.
Since the cost function in \eqref{eq:MainUpdata:WPM:Open}
is differentiable and strongly convex,
we apply the accelerated projection method with fixed momentum
to solve it \cite{nesterov2018lectures,hong2022adapting}.
To avoid using a line search or computing the exact Lipschitz constant,
which would increase the computational cost, we instead estimate an upper bound.
The Hessian matrix of the cost function in \eqref{eq:MainUpdata:WPM:Open}
is $\umB_k + \stepsize_k \umA^\HTrans \umA$.
\MRcb{Clearly, we have the following relations
\begin{equation}
\label{eq:SmallLargeEigRelations}
 \begin{array}{rcl}
 	 \mu_{\min}(\umB_k+\umA^\HTrans\umA)
    &\geq & \mu_{\min}(\umB_k) + \mu_{\min}(\umA^\HTrans\umA) ,\\
     \mu_{\max}(\umB_k+\umA^\HTrans\umA) &\leq & \mu_{\max}(\umB_k) + \mu_{\max}(\umA^\HTrans\umA),
 \end{array}	
\end{equation}
where $\mu_{\min}(\cdot)$ and $\mu_{\max}(\cdot)$
denote the smallest and largest eigenvalues, respectively.
Let $L_{\umA}$ denote the largest eigenvalue of $\umA^\HTrans \umA$.
In CS MRI, the kspace data is under-sampled
so that the smallest eigenvalue of $\umA^\HTrans \umA$ is zero.
Using the relations in \eqref{eq:SmallLargeEigRelations}
and $\umB_k=\tau_k^{-1} \umI_N - (\uvu_k \uvu_k^\HTrans)/(\rho_k^{\umB})$
(cf. \Cref{alg:ModfiedZero:SR1}),
the largest and smallest eigenvalues of
$\umB_k+\stepsize_k \umA^\HTrans \umA$
are upper and lower bounded by $L_k = \tau_k^{-1}+\stepsize_k L_{\umA}$
and
$\sigma_k = \tau_k^{-1} - \uvu_k^\HTrans \uvu_k / \rho_k^{\umB}$,
respectively,
where
\Cref{lemma:boundedHessian} shows that $\sigma_k > 0$}.
\Cref{alg:ComputeWPM} describes
the accelerated projection method with fixed momentum
for solving \eqref{eq:MainUpdata:WPM:Open}.

 

\begin{algorithm}[t]        
\caption{Modified Memory Efficient Self-Scaling Hermitian Rank-1 Method}
\label{alg:ModfiedZero:SR1} 
\begin{algorithmic}[1]
\REQUIRE $\uvx_{k-1}$, $\uvx_k$, $\nabla f(\uvx_{k-1})$, $\nabla f(\uvx_k)$, \MRcb{$\delta=10^{-8}$},
\\
\quad \quad \quad \quad \quad$\theta_1\in (0,1)$, and $\theta_2\in(1,\infty)$
\STATE Set $\uvs_k\leftarrow \uvx_k-\uvx_{k-1}$ and $\uvm_k\leftarrow (\nabla f(\uvx_k)-\nabla f(\uvx_{k-1}))$
\STATE Compute $\beta$ such that 
\begin{equation}
\label{eq:alg:Zero:SC:SR1:alpha}
\begin{array}{c}
     \min_\beta\{\beta\in[0,1]|\uvv_k=\beta\uvs_k+(1-\beta)\uvm_k\}  \\
     \text{satisfies}~\theta_1\leq \frac{\Re(\langle\uvs_k,\uvv_k\rangle)}{\langle\uvs_k,\uvs_k\rangle}~\text{and}~\frac{\langle\uvv_k,\uvv_k\rangle}{\Re(\langle\uvs_k,\uvv_k\rangle)}\leq\theta_2 
\end{array}    
\end{equation} \label{alg:ModfiedZero:SR1:stepv_k}
\STATE Compute $\tau_k \leftarrow \frac{\langle\uvs_k,\uvs_k\rangle}{\Re(\langle\uvs_k,\uvv_k\rangle)}-\sqrt{\left(\frac{\langle\uvs_k,\uvs_k\rangle}{\Re(\langle\uvs_k,\uvv_k\rangle)}\right)^2-\frac{\langle\uvs_k,\uvs_k\rangle}{\langle\uvv_k,\uvv_k\rangle}}$\label{alg:ModfiedZero:SR1:steptau_k}\\[5pt]
\STATE $\rho_k\leftarrow \Re(\langle \uvs_k-\tau_k\uvv_k,\uvv_k\rangle)$\label{alg:ModfiedZero:SR1:stepInnerrhok}
\IF{$\rho_k \leq \delta \|\uvs_k-\tau_k\uvv_k\|\|\uvv_k\|$ \label{alg:ModfiedZero:SR1:u_k0}}
\STATE $\uvu_k\leftarrow \bm 0$
\ELSE
\STATE $\uvu_k\leftarrow \uvs_k-\tau_k\uvv_k$
\ENDIF
\STATE $\rho_k^{\umB} \leftarrow \tau_k^2 \rho_k + \tau_k \uvu_k^\HTrans \uvu_k$
\STATE {\bf Return:} $\umH_k \leftarrow \tau_k\umI_N+\frac{\uvu_k\uvu_k^\HTrans}{\rho_k}$ and $
\umB_k \leftarrow \tau_k^{-1} \umI_N - \frac{\uvu_k \uvu_k^\HTrans}{\rho_k^{\umB}}
$
\end{algorithmic}
\end{algorithm}

\begin{algorithm}[t]        
\caption{Accelerated Projection Method with Fixed Momentum for Computing the WPM}    
\label{alg:ComputeWPM} 
\begin{algorithmic}[1]
\REQUIRE $\bar{\uvx}_1=\uvx_k$, $\uvz_1=\uvx_k$, \MRcb{tolerance $\epsilon=10^{-6}$}, stepsize $\frac{1}{L_k}$ and $\kappa_k=\frac{L_k}{\sigma_k}$ \MRcb{where $L_k$ (respectively, $\sigma_k$) denotes the upper (respectively, lower) bound of the largest (respectively, smallest) eignenvalue of $\umB_k+\stepsize_k\umA^\HTrans\umA$}
\ENSURE 
\FOR {$i=1,2,\dots$}
\STATE $\bar{\uvx}_{i+1}\leftarrow \proxW{\iota_\mathcal C}{\umI}{(\uvz_i-\frac{1}{L_k} \nabla G_k(\uvz_i) )}$\label{alg:ComputeWPM:Project}
\IF{$\|\bar{\uvx}_{i+1}-\bar{\uvx}_i\|\leq \epsilon$}
\STATE break
\ELSE
\STATE $\uvz_{i+1}\leftarrow \bar{\uvx}_{i+1}+\frac{\sqrt{\kappa_k}-1}{\sqrt{\kappa_k}+1}(\bar{\uvx}_{i+1}-\bar{\uvx}_i)$
\ENDIF
\ENDFOR
\end{algorithmic}
\end{algorithm}


 
\section{Convergence Analysis}
\label{sec:convAnalysis}

This section provides the convergence analysis of \Cref{alg:PropMethod}
for solving \eqref{eq:CSMRIReco:origP:GradDenoiser}.
Before presenting the convergence results,
we first introduce two assumptions and then provide four lemmas
that simplify presenting the convergence proof.

%
\begin{assume}[$L$-Smooth regularizer]
\label{assum:LipGrad}
We assume the gradient of $f(\uvx)$ is $L$-Lipschitz continuous,
meaning that for all $\uvx_1,\,\uvx_2\in\mathbb C^N$,
there exists a positive constant $L$
such that the following inequality holds:
\begin{equation}
\label{eq:gradLip:l}
\|\nabla f(\uvx_1)-\nabla f(\uvx_2)\|\leq L \, \|\uvx_1-\uvx_2\|.
\end{equation}
\end{assume}

\begin{assume}[Constrained proximal Polyak-\L{}ojasiewicz inequality
\cite{karimi2016linear,wang2019stochastic}]
\label{assum:PLCond}
Define 
\begin{equation}
\label{eq:def:Suggorate_for_D_h}
\begin{array}{l}
\mathcal S_{h}(\bar{\uvx},\uvx,\uvg,\umW,\stepsize)
\defequ \Re\big\{\langle \uvg,\bar{\uvx}-\uvx \rangle\big\}
+\frac{1}{2\stepsize}\|\bar{\uvx}-\uvx\|_{\umW}^2
\\[3pt]
\quad\quad\quad\quad\quad\quad\quad\quad\quad\quad\quad\quad
+h(\bar{\uvx})-h(\uvx),
\end{array}
\end{equation}
and
\begin{equation}
\label{eq:def:D_h}
\mathcal D_{h}^{\mathcal C}(\uvx,\uvg,\umW,\stepsize)
\defequ -\frac{2}{\stepsize} \min_{\bar{\uvx}\in \mathcal C}
\mathcal S_{h}(\bar{\uvx},\uvx,\uvg,\umW,\stepsize),
\end{equation}
where $\umW\in\mathbb C^{N\times N},~\umW\succ0$
is a Hermitian positive definite matrix,
$h$ was defined in \eqref{eq:CSMRIReco:origP:GradDenoiser}
and $\stepsize$ is a positive constant.
If there exists a positive constant $\nu$
such that the following inequality holds:
\begin{equation}
\label{eq:PLIneq}
\mathcal D_{h}^{\mathcal C}(\uvx,\nabla f(\uvx),\umI_N,\stepsize)
\geq 2\nu (F(\uvx)-F^*), \ \forall \uvx \in \mathcal C,
\end{equation}
then we say $F(\uvx)$ satisfies the constrained proximal Polyak-\L{}ojasiewicz inequality.
$F^*$ denotes the optimal value of \eqref{eq:CSMRIReco:origP:GradDenoiser}.
\end{assume}

\MRcb{\Cref{assum:LipGrad} is a moderate assumption and is commonly used in convergence analysis for differentiable functions, see \cite{cohen2021has,hurault2021gradient}.  We analysed the validity of \Cref{assum:LipGrad} for the used network architecture (see \Cref{fig:NNArchitecture}), along with empirical validation in the supplementary material.
 \Cref{assum:PLCond} covers certain nonconvex properties of $F(\uvx)$ and is commonly used in nonconvex analysis~\cite{karimi2016linear}.} 
Next we introduce \Cref{lemma:DescentLemma,lemma:D_h:GM:inequality,lemma:Property:D_h:stepsize,lemma:boundedHessian}
which are useful for the following convergence proofs.


\begin{lemma}[Majorizer of $f$]
\label{lemma:DescentLemma}
    Let $f:\,\mathbb C^N\rightarrow (-\infty,\infty]$ be an $L$-smooth function ($L>0$). Then for any $\uvx_1,
\uvx_2\in\mathbb C^N$, we have
\begin{equation}
    \label{eq:DescentLemma}
    f(\uvx_2)\leq f(\uvx_1)+\Re\big\{\langle \nabla f(\uvx_1),\uvx_2-\uvx_1\rangle\big\}+\frac{L}{2}\|\uvx_1-\uvx_2\|_2^2.
\end{equation}
\end{lemma}

\begin{lemma}
\label{lemma:D_h:GM:inequality}
For any fixed $\umW\in\mathbb C^{N\times N},~\umW\succ0$ and $\stepsize>0$,
we have the following inequality
$$
\mathcal D_h^\mathcal C(\uvx,\nabla f(\uvx),\umW,\stepsize)
\geq \|\mathcal G_{\frac{1}{\stepsize},\umW}^{f,h}(\uvx)\|_{\umW}^2,
\quad \forall \uvx\in\mathcal C,
$$
where
$\mathcal G_{\frac{1}{\stepsize},\umW}^{f,h}(\uvx)
\defequ \frac{1}{\stepsize}(\uvx-\bar{\uvx}^+)$
with   
\begin{equation}
\label{eq:OptimalSoluS_h}
\bar{\uvx}^+
\defequ
\arg\min\limits_{\bar{\uvx}\in\mathcal C}
\mathcal S_h(\bar{\uvx},\uvx,\nabla f(\uvx),\umW,\stepsize).
\end{equation}
\end{lemma}

\begin{lemma}
\label{lemma:Property:D_h:stepsize}
For any fixed $\umW\in\mathbb C^{N\times N},~\umW\succ 0$,
a differentiable function $f$ and a convex function $h$,
$\mathcal D_h^\mathcal C(\uvx,\nabla f(\uvx),\umW,\stepsize)$
satisfies the following inequality
for $\forall \stepsize_1\geq\stepsize_2>0$:
\begin{equation}
\label{eq:D_h:stepsize:inequality}
\mathcal D_h^\mathcal C(\uvx,\nabla f(\uvx),\umW,\stepsize_2)
\geq
\mathcal D_h^\mathcal C(\uvx,\nabla f(\uvx),\umW,\stepsize_1). 
\end{equation}
\end{lemma}



\begin{lemma}[Bounded Hessian]
\label{lemma:boundedHessian}
The \MRcb{approximate} Hessian matrices generated by \Cref{alg:ModfiedZero:SR1}
satisfy the following inequality
$$
\underline{\eta}\,\umI \preceq \umB_k \preceq \overline{\eta}\,\umI,
$$
where $0<\underline{\eta}<\overline{\eta}<\infty$.
\end{lemma}
The proof of \Cref{lemma:DescentLemma} is similar to the one in the real-valued case \cite{nesterov2018lectures}.
The only difference is the use of
$\Re\big\{\langle \nabla f(\uvx_1),\uvx_2-\uvx_1\rangle\big\}$
instead of $\langle \nabla f(\uvx_1),\uvx_2-\uvx_1\rangle$
to account for the complex domain.
Therefore, we omit the proof here for brevity.
\Cref{proof:lemma:D_h:GM:inequality,proof:lemma:Property:D_h:stepsize,proof:lemma:boundedHessian}
give the proofs of 
\Cref{lemma:D_h:GM:inequality,lemma:Property:D_h:stepsize,lemma:boundedHessian}.
With these in place, we are able to derive our main convergence results
in \Cref{them:ConvResults}.

\begin{theorem}[Convergence results]
\label{them:ConvResults}
Applying \Cref{alg:PropMethod}
to solve  \eqref{eq:CSMRIReco:origP:GradDenoiser},
we can establish the following convergence results:
\begin{itemize}
\item
Let $\stepsize_k\leq \frac{\underline{\eta}}{L}$,
and define $\Delta_K \defequ \min_{k\leq K} \|\uvx_{k+1}-\uvx_k\|_2^2$.
Under \Cref{assum:LipGrad}, by running \Cref{alg:PropMethod} $K$ iterations
to solve \eqref{eq:CSMRIReco:origP:GradDenoiser},
we have
$$
\Delta_K \leq \frac{2\left(F(\uvx_1)-F^*\right)}{LK},
$$ 
where $F^*$ denotes the minimum of \eqref{eq:CSMRIReco:origP:GradDenoiser}
and $\uvx_1$ is the initial iterate. 
 
\item
Let $\stepsize_k
\leq \min\left\{\frac{\overline{\eta}}{\nu},\frac{\underline{\eta}}{L}\right\}$.
Under \Cref{assum:LipGrad,assum:PLCond},
we have the following convergence rate bound for the cost value
$$
F(\uvx_{k+1})-F^*
\leq \Big(1-\frac{\nu\,\stepsize_{\mathrm{min}}}{\overline{\eta}}\Big)^k \,
\big (F(\uvx_1)-F^*\big),
$$
where $\stepsize_{\mathrm{min}}=\min_k \{\stepsize_k\}$.

\item
Let $\stepsize_k\leq \frac{\underline{\eta}}{L}$.
Choose $k \in \{1,\ldots,K-1\}$ uniformly at random.
Then under \Cref{assum:LipGrad,assum:PLCond},
we have the following convergence rate bound in expectation
$$
\mathbb{E}\Big[F(\uvx_{k})-F^*\Big]
\leq \frac{\overline{\eta}\left(F(\uvx_1)-F^*\right)}{\nu\stepsize_{\mathrm{min}}K}.
$$
\end{itemize}
\end{theorem}
\Cref{proof:them:ConvResults}
presents the proof.
The first part of \Cref{them:ConvResults} shows that
$\Delta_K\rightarrow 0$ as $K\rightarrow \infty$.
Combining with the summation in \eqref{eq:sum:gamma_k:Cost:inequality}
implies convergence to a fixed point.
The second part establishes that
the cost function sequence
converges linearly to the minimal cost value.
The third part demonstrates that if one selects a random iterate,
then the cost value converges sublinearly in expectation.
In this paper, we simply choose the last iterate as the output.
However,
\Cref{sec:numericalExp:spiral}
discusses the third term experimentally to validate our analysis. \MRcb{Note that \Cref{assum:PLCond} is a special case of the Kurdyka-{\L}ojasiewicz (KL) inequality with exponent $1/2$~\cite{attouch2010proximal}. While our analysis establishes convergence rate under the PL inequality, it could potentially be extended to the more general KL inequality, see~\cite{hong2025GKrylov}. Moreover, if the Dennis-Mor\'e condition~\cite{dennis1977quasi} holds, one can expect a superlinear convergence rate. However, a full study of this direction is beyond the scope of the present paper and is left for future work.}

\section{Numerical Experiments}
\label{sec:numericalExp}

\begin{figure}[t]
\centering
\subfigure[Brain]{
\includegraphics[width=0.2\textwidth]{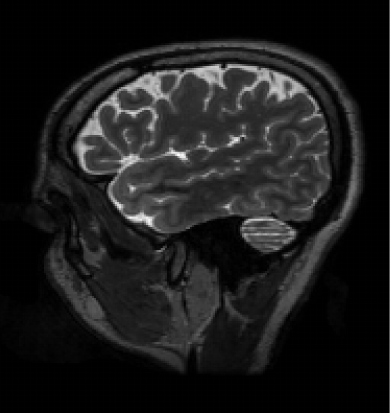}
}	
\subfigure[Knee]{
\includegraphics[width=0.2\textwidth]{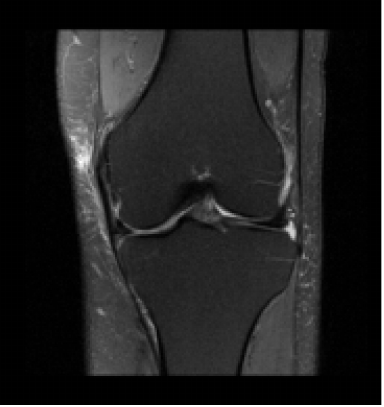}
}
\caption{Magnitude of the complex-valued ground-truth images.}
\label{fig:GT}
\end{figure}

\begin{figure}[t]
\centering
\subfigure[Spiral]{
	\includegraphics[width=0.2\textwidth]{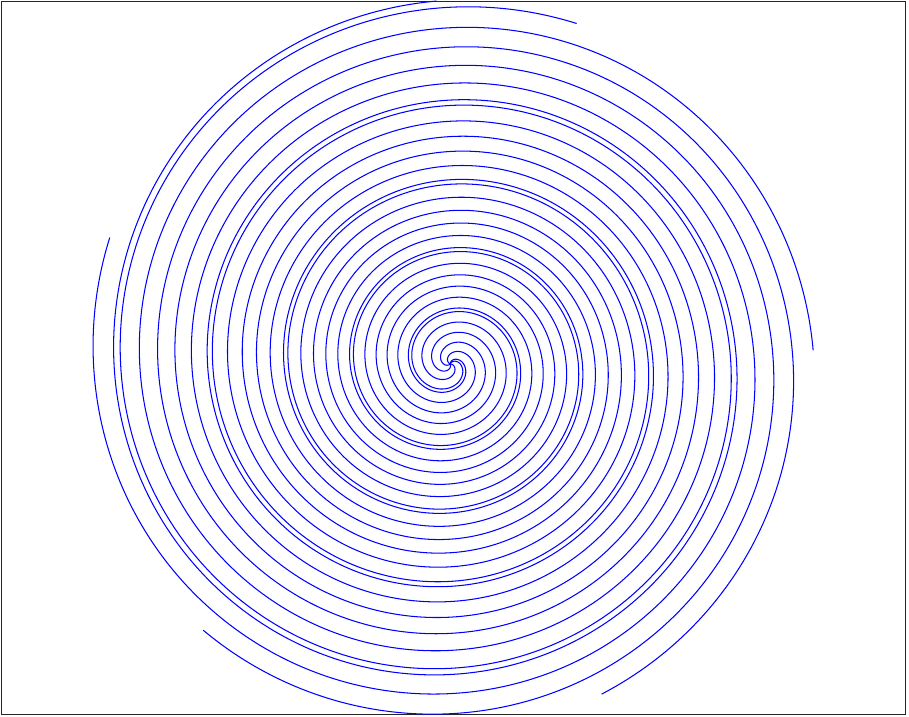}}
	\subfigure[Radial]{
	\includegraphics[width=0.2\textwidth]{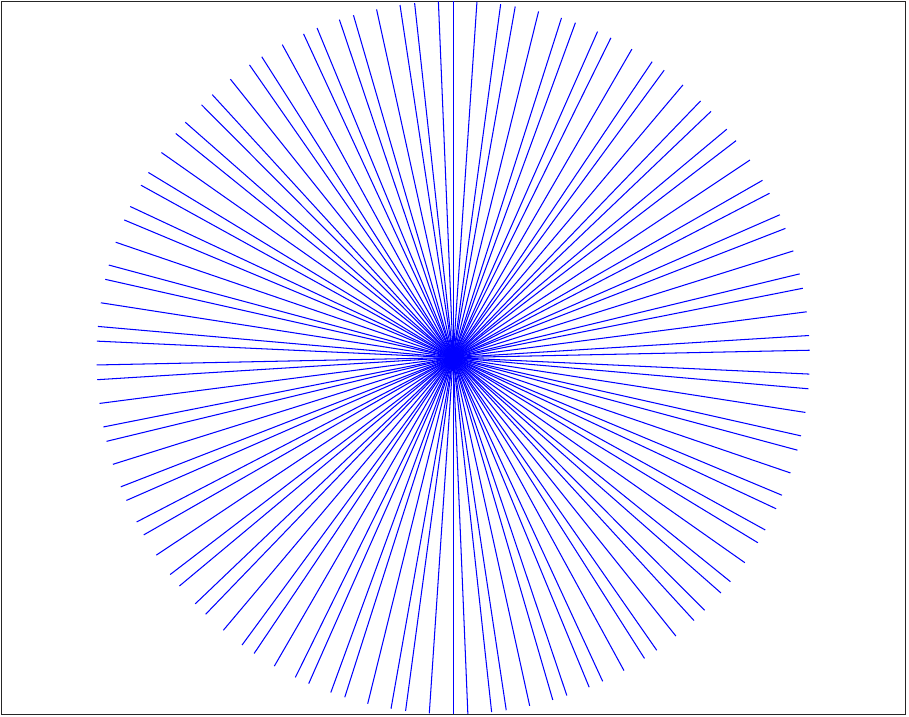}}
	
	\subfigure[Cartesian Sampling Mask]{
	\includegraphics[width=0.2\textwidth]{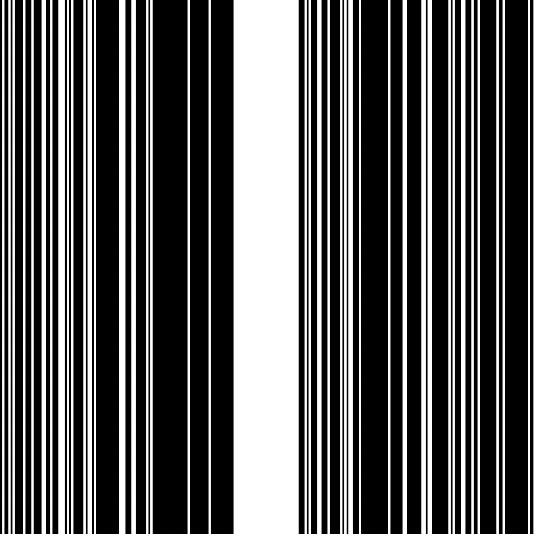}}
	\caption{The spiral (a) and radial (b) sampling trajectories
    and the Cartesian sampling mask (c) used in the experiments.}
	\label{fig:TrjsMask}
\end{figure}

\begin{figure*}[t]
\centering
\includegraphics[width=0.8\textwidth]{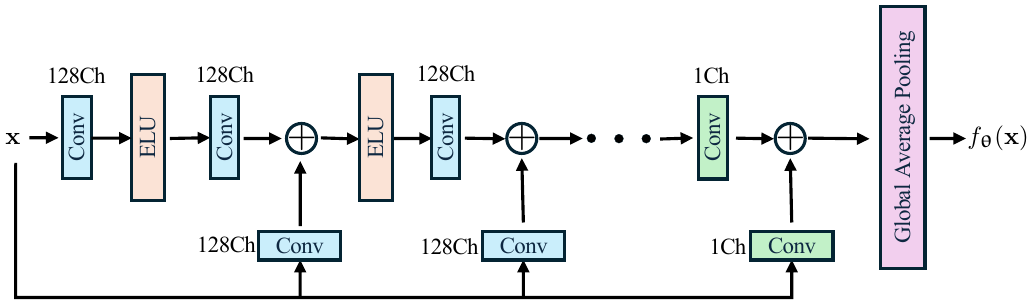}
\caption{\MRcb{The neural network architecture used
for the energy function $f_{\vtheta}(\uvx)$,
based on \cite{cohen2021has}.
The convolutional kernel size is $3\times 3$ with stride $1$.}}
\label{fig:NNArchitecture}	
\end{figure*}

This section investigates the performance of the proposed method
for CS MRI reconstruction using spiral, radial, and Cartesian k-space sampling patterns. We first describe the experimental and algorithmic settings.
Then, we present the reconstruction results
and experimentally demonstrate the convergence of the proposed method,
validating our theoretical analysis.

%
%
%
{\noindent \bf  Experimental Settings:}
We evaluated the performance of the proposed method using brain and knee MRI datasets.
For the brain images, we used the dataset from \cite{aggarwal2018modl},
which consists of $360$ images for training and $164$ images for testing.
For knee images, we employed the NYU fastMRI \cite{zbontar2018fastmri} multi-coil knee dataset.
The ESPIRiT algorithm  \cite{uecker2014espirit} was first applied
to reconstruct complex-valued images from multi-coil k-space data.
All brain and knee images were cropped and resized to a resolution of $256\times 256$
and \MRcb{normalized such that their maximum magnitude is one.}
We employed the network architecture proposed in \cite{cohen2021has} to construct $f(\uvx)$.
Unlike \cite{cohen2021has}, we included bias terms in the network,
as this provided improved performance in our experiments.
\MRcb{For completeness, \Cref{fig:NNArchitecture} presents the used network architecture.
The number of layers is set to six, which is identical to that in \cite{cohen2021has}.}
Noisy images were generated by adding independent and identically distributed (i.i.d.)
Gaussian noise with a variance of $1/255$.
\MRcb{The network was trained to enforce $\uvx-\nabla f(\uvx)$ to be a denoiser
and $\nabla f(\uvx)$ was computed with PyTorch's autograd function,
i.e., \texttt{torch.autograd.grad}.}
The training process used the mean squared error as the loss function,
with a batch size of $64$.
Optimization was performed using the ADAM algorithm \cite{kingma2014adam}
with a learning rate $10^{-3}$ over a total of $18,000$ iterations.
Additionally, we halved the learning rate at each $4,000$ iterations.
Although we trained distinct denoisers for brain and knee images,
the same denoiser was employed across different sampling acquisitions.
\MRcb{Note that training denoisers on larger datasets may lead to improved denoisers
and then better reconstruction.
However, the main focus of this paper is to investigate
numerical efficiency of gradient-driven denoisers based CS MRI reconstruction
with convergence guarantees.
Thus, we leave the exploration of this direction to future work.
}

To evaluate reconstruction performance, we selected $6$ brain and $6$ knee images from the test datasets as the ground-truth. \Cref{fig:GT} presents the magnitudes of two of these twelve complex-valued ground truth images. Due to space limitations, the remaining ten brain and knee ground-truth images are presented in the supplementary material. For the spiral trajectory, we used $6$ interleaves, $1688$ readout points, and $32$ coils, whereas the radial trajectory involved $55$ spokes with golden-angle rotation, $1024$ readout points, and $32$ coils. \Cref{fig:TrjsMask} illustrates the sampling trajectories and mask used in our experiments. To generate the k-space data, we first applied the forward model with the corresponding trajectories and sensitivity maps to the ground-truth images. We then added complex i.i.d. Gaussian noise with zero mean and a variance of $10^{-4}$, resulting in noisy measurements with an input SNR of approximately $21$dB. In the reconstruction step, we used coil compression \cite{zhang2013coil} to reduce the $32$ coils to $12$ virtual coils in order to reduce computational complexity. \Cref{sec:numericalExp:spiral,sec:numericalExp:radial,sec:numericalExp:Cartesian} specifically analyze one brain and one knee test image acquired using the spiral, radial, and Cartesian sampling trajectories. Additional results on other test images are provided in the supplementary material. All experiments were implemented using PyTorch and executed on an NVIDIA A100 GPU.

{\noindent \bf Algorithmic Settings:} 
We primarily compared our method with the projected gradient descent method
(dubbed GD) \cite{cohen2021has}
and the proximal gradient method (dubbed PG) \cite{hurault2021gradient},
as both methods also come with convergence guarantees, similar to ours.
In addition, we compared with the accelerated proximal gradient method
(dubbed APG)~\cite{li2015accelerated}
that also provides convergence guarantees for solving nonconvex problems
such as \eqref{eq:CSMRIReco:origP:GradDenoiser}.
Since we normalized the maximum magnitude of all images to one,
we set $\mathcal{C} = \{\uvx \mid \|\uvx\|_\infty \leq 1\}$
and all methods used this constraint.
\MRcb{Although such a constraint is not needed in practical CS MRI reconstruction,
we considered it here primarily to test the applicability of CQNPM to constrained problems.
Following \cite{hurault2022proximal}, which was also adopted in PnP-LBFGS,
we attempted to train a network to meet the requirements of PnP-LBFGS for CS MRI reconstruction.
However, we observed that PnP-LBFGS diverged.
Therefore, we compared our method against PnP-LBFGS on the image deblurring problem
using the same experimental setup in their open-source code.
The results, provided in the supplementary material,
show that our method converged faster than PnP-LBFGS in terms of PSNR
with respect to both iterations and wall time.
}

For plots involving $F^*$,
we ran APG for $500$ iterations
and set
\MRcb{$
F^* = F(\uvx_{500}).
$ The stepsize $\stepsize_k$ in \Cref{alg:PropMethod} is set to be one which we found it works well for our experimental settings. Alternatively, one can choose $\stepsize_k$ with a backtracking line search strategy to satisfy \eqref{eq:proveConv:descent:F}.}
\Cref{alg:ModfiedZero:SR1} used
$\delta = 10^{-8}, \theta_1 = 2\times 10^{-6}, \theta_2 = 200$.
\MRcb{The parameter $\delta$ is commonly used in quasi-Newton methods
to guard numerical stability when $\uvv_k$ and $\uvs_k$ are closely related.
Parameters $\theta_1$ and $\theta_2$
control the Hermitian positive definiteness of the estimated Hessian matrices
but still allow greater flexibility of estimation.
In view of the proof in \Cref{lemma:boundedHessian}, we have 
$
\frac{1}{2\theta_2}\umI_N \preceq \umH_k \preceq \left(\frac{1+\delta}{\delta\theta_1}\right)\umI_N.
$
So $\theta_1$ and $\theta_2$ should be positive,
with $\theta_1$ chosen small and $\theta_2$ chosen large.
In \Cref{alg:ComputeWPM}, we always used the previous iterate as the new initial value.
Therefore, we found that setting the maximum number of iterations to $15$
and $\varepsilon = 10^{-6}$ worked well in practice.
}


\subsection{Spiral Acquisition Reconstruction}
\label{sec:numericalExp:spiral}

\begin{figure}[t]
	\centering
	\begin{tikzpicture}
  \pgfmathsetmacro{\minybrainOne}{-13.73+0.3e-4} 
  \begin{axis}[
      name=CostIter,
     at={(0,0)},
    anchor=south west,
	width=0.28\textwidth,
 	xmin=0, xmax=150,
      xlabel={Iteration},
      ylabel={\MRcb{$F(\uvx_k)-F^*+\varepsilon$}},
      tick label style={font=\fontsize{5}{5.5}\selectfont},
      label style={font=\fontsize{7}{8.4}\selectfont},
      ylabel style = {yshift=-5pt},
      xlabel style = {yshift=2pt},
      grid=both,
      ymode=log,
      legend style={
  at={(0.5,1.03),
  font=\fontsize{7}{8.4}\selectfont},
  anchor=north west,
  draw=none,          
  fill=none,          
  cells={align=left},  
  }
  ]
    
      \addplot [black,dotted,line width=1pt] table [y expr=\thisrowno{1}-\minybrainOne, x expr=\coordindex] {./figs/MRI/Spiral/DeepBrain1/REDEnergy_GD/lst_cost_time.txt};
    \addlegendentry{GD}
    
     \addplot [green,solid,line width=1pt] table [y expr=\thisrowno{1}-\minybrainOne, x expr=\coordindex] {./figs/MRI/Spiral/DeepBrain1/REDEnergy_ISTA/lst_cost_time.txt};
     \addlegendentry{PG}

    \addplot [blue,dash dot,line width=1pt] table [x expr=\coordindex,y expr=\thisrowno{1}-\minybrainOne] {./figs/MRI/Spiral/DeepBrain1/REDEnergy_FISTA_LL/lst_cost_time.txt};
     \addlegendentry{APG}

    \addplot [red,dashed,line width=1pt] table [x expr=\coordindex, y expr=\thisrowno{1}-\minybrainOne] {./figs/MRI/Spiral/DeepBrain1/REDEnergy_QNP/lst_cost_time.txt};
    \addlegendentry{CQNPM}

  \end{axis}
  \node[anchor=north west, xshift= 42pt, yshift=-18pt] at (CostIter.south west) {(a)};

   \begin{axis}[
      name=CostTime,
        at={(CostIter.south east)},
    anchor=south west,
	width=0.28\textwidth,
	xshift=10pt,
 xtick={0, 15, 30, 45},
 xticklabels={0, 15, 30, 45},
yticklabels={},
 	xmin=0, xmax=45,
      xlabel={Wall Time (Seconds)},
       tick label style={font=\fontsize{5}{5.5}\selectfont},
      label style={font=\fontsize{7}{8.4}\selectfont},
      ylabel style = {yshift=-5pt},
      xlabel style = {yshift=2pt},
      grid=both,
      ymode=log,
      legend style={
  at={(0.67,1),
  font=\fontsize{5}{5.5}\selectfont},
  anchor=north west,
  draw=none,          
  fill=none,          
  cells={align=left},  
  }
  ]
    
    \addplot [red,dashed,line width=1pt] table [x index=0,
    y expr=\thisrowno{1} - \minybrainOne] {./figs/MRI/Spiral/DeepBrain1/REDEnergy_QNP/lst_cost_time.txt};
    
    \addplot [blue,dash dot,line width=1pt] table [x index=0,y expr=\thisrowno{1}-\minybrainOne] {./figs/MRI/Spiral/DeepBrain1/REDEnergy_FISTA_LL/lst_cost_time.txt};

    \addplot [green,solid,line width=1pt] table [y expr=\thisrowno{1}-\minybrainOne, x index=0] {./figs/MRI/Spiral/DeepBrain1/REDEnergy_ISTA/lst_cost_time.txt};
  
     \addplot [black,dotted,line width=1pt] table [y expr=\thisrowno{1}-\minybrainOne, x index=0] {./figs/MRI/Spiral/DeepBrain1/REDEnergy_GD/lst_cost_time.txt};

  \end{axis}
  \node[anchor=north west, xshift= 42pt, yshift=-18pt] at (CostTime.south west) {(b)};
  
\begin{axis}[
      name=PSNRIter,
     at={(CostIter.south)},
    anchor=north,
	width=0.28\textwidth,
	yshift = -36pt,
 xtick={0, 50, 100, 150},
  xticklabels={0, 50, 100, 150},
   ytick={20, 28, 35, 41},
  yticklabels={20, 28, 35, 41},
  	xmin=0, xmax=150,
    ymin=19, ymax=41,
      xlabel={Iteration},
      ylabel={PSNR (dB)},
       tick label style={font=\fontsize{5}{5.5}\selectfont},
      label style={font=\fontsize{7}{8.4}\selectfont},
      grid=both,
      legend style={
  at={(0.67,0.36),
  font=\fontsize{5}{5.5}\selectfont},
  anchor=north west,
  draw=none,          
  fill=none,          
  cells={align=left},  
  }
  ]
    
    \addplot [red,dashed,line width=1pt] table [y index=1, x expr=\coordindex] {./figs/MRI/Spiral/DeepBrain1/REDEnergy_QNP/lst_psnr_time.txt};

    \addplot [blue,dash dot,line width=1pt] table [y index=1, x expr=\coordindex] {./figs/MRI/Spiral/DeepBrain1/REDEnergy_FISTA_LL/lst_psnr_time.txt};

    \addplot [green,solid,line width=1pt] table [y index=1, x expr=\coordindex] {./figs/MRI/Spiral/DeepBrain1/REDEnergy_ISTA/lst_psnr_time.txt};

     \addplot [black,dotted,line width=1pt] table [y index=1, x expr=\coordindex] {./figs/MRI/Spiral/DeepBrain1/REDEnergy_GD/lst_psnr_time.txt};

  \end{axis}
  \node[anchor=north west, xshift= 42pt, yshift=-18pt] at (PSNRIter.south west) {(c)};

  \begin{axis}[
      name=PSNRTime,
     at={(CostTime.south)},
    anchor=north,
    yshift = -36pt,
	width=0.28\textwidth,
  xtick={0, 15, 30, 45},
 xticklabels={0, 15, 30, 45},
  yticklabels={},
  	xmin=0, xmax=45,
    ymin=19, ymax=41,
     xlabel={Wall Time (Seconds)},
      tick label style={font=\fontsize{5}{5.5}\selectfont},
      label style={font=\fontsize{7}{8.4}\selectfont},
      grid=both,
      legend style={
  at={(0.67,0.36),
  font=\fontsize{5}{5.5}\selectfont},
  anchor=north west,
  draw=none,          
  fill=none,          
  cells={align=left},  
  }
  ]
    
    \addplot [red,dashed,line width=1pt] table [y index=1, x index=0] {./figs/MRI/Spiral/DeepBrain1/REDEnergy_QNP/lst_psnr_time.txt};

    \addplot [blue,dash dot,line width=1pt] table [y index=1, x index=0] {./figs/MRI/Spiral/DeepBrain1/REDEnergy_FISTA_LL/lst_psnr_time.txt};
    
    \addplot [green,solid,line width=1pt] table [y index=1, x index=0] {./figs/MRI/Spiral/DeepBrain1/REDEnergy_ISTA/lst_psnr_time.txt};
     
     \addplot [black,dotted,line width=1pt] table [y index=1, x index=0] {./figs/MRI/Spiral/DeepBrain1/REDEnergy_GD/lst_psnr_time.txt};

  \end{axis}
  \node[anchor=north west, xshift= 42pt, yshift=-18pt] at (PSNRTime.south west) {(d)};
\end{tikzpicture}
\caption{Comparison of different methods with spiral acquisition on the brain image
for $\varepsilon = 4\times10^{-3}$.
(a), (b): cost values versus iteration and wall time;
(c), (d): PSNR values versus iteration and wall time. }
\label{fig:SpiralBrain1CostPSNR}
\end{figure}
\input{figs/SpiralBrain1RecoIm}
\Cref{fig:SpiralBrain1CostPSNR} presents the cost and PSNR values of each method with respect to the number of iterations and wall time for the reconstruction with spiral acquisition on the brain image. Fig.s~\ref{fig:SpiralBrain1CostPSNR}~(a) and (c) show that our method converges faster than the others, reaching a lower cost value and higher PSNR at the same number of iterations. Moreover, we observed that APG is faster than PG, and PG is faster than GD in terms of the number of iterations, which aligns with our expectations. From Fig.s~\ref{fig:SpiralBrain1CostPSNR}~(b) and (d), we saw that our method is also the fastest algorithm in terms of wall time. Moreover, we observed that GD becomes faster than PG in terms of wall time because PG needs to solve a proximal mapping  requiring to execute $\umA\uvx$ multiple times.  However, APG is still faster than GD even it also requires to solve a proximal mapping.

\Cref{fig:SpiralBrain1:visual}
shows the reconstructed images at the $50$th and $100$th iterations.
Our method recovered significantly clearer images than other methods
at the same number of iterations.
Moreover, the corresponding error maps clearly show that
our method yielded small reconstruction errors.
The supplementary material presents the reconstruction results of the knee image
using the same spiral acquisition, as well as results for other additional test images,
where similar trends were observed across all methods.

\subsection{Radial Acquisition Reconstruction}
\label{sec:numericalExp:radial}

\begin{figure}[t]
	\centering
	\begin{tikzpicture}
  \pgfmathsetmacro{\minybrainOne}{0.01+5e-5}
  \begin{axis}[
      name=CostIter,
     at={(0,0)},
    anchor=south west,
	width=0.28\textwidth,
 xtick={0, 30, 60, 90,120},
  xticklabels={0, 30, 60, 90,120},
 	xmin=0, xmax=120,
      xlabel={Iteration},
      ylabel={\MRcb{$F(\uvx_k)-F^*+\varepsilon$}},
      tick label style={font=\fontsize{5}{5.5}\selectfont},
      label style={font=\fontsize{7}{8.4}\selectfont},
      ylabel style = {yshift=-5pt},
      xlabel style = {yshift=2pt},
      grid=both,
      ymode=log,
      legend style={
  at={(0.55,1.03),
  font=\fontsize{6}{7.2}\selectfont},
  anchor=north west,
  draw=none,          
  fill=none,          
  cells={align=left},  
  }
  ]
    
       \addplot [black,dotted,line width=1pt] table [y expr=\thisrowno{1}-\minybrainOne, x expr=\coordindex] {./figs/MRI/Radial/Knee1/REDEnergy_GD/lst_cost_time.txt};
    \addlegendentry{GD}

    \addplot [green,solid,line width=1pt] table [y expr=\thisrowno{1}-\minybrainOne, x expr=\coordindex] {./figs/MRI/Radial/Knee1/REDEnergy_ISTA/lst_cost_time.txt};
     \addlegendentry{PG}

         \addplot [blue,dash dot,line width=1pt] table [x expr=\coordindex,y expr=\thisrowno{1}-\minybrainOne] {./figs/MRI/Radial/Knee1/REDEnergy_FISTA_LL/lst_cost_time.txt};
     \addlegendentry{APG}
    
    \addplot [red,dashed,line width=1pt] table [x expr=\coordindex, y expr=\thisrowno{1}-\minybrainOne] {./figs/MRI/Radial/Knee1/REDEnergy_QNP/lst_cost_time.txt};
    \addlegendentry{CQNPM}
  
  \end{axis}
  \node[anchor=north west, xshift= 42pt, yshift=-18pt] at (CostIter.south west) {(a)};

   \begin{axis}[
      name=CostTime,
        at={(CostIter.south east)},
    anchor=south west,
	width=0.28\textwidth,
	xshift=10pt,
 xtick={0, 15, 30, 45},
  xticklabels={0, 15, 30, 45},
ytick={0.1, 1, 100},
yticklabels={},
 	xmin=0, xmax=45,
      xlabel={Wall Time (Seconds)},
      tick label style={font=\fontsize{5}{5.5}\selectfont},
      label style={font=\fontsize{7}{8.4}\selectfont},
      ylabel style = {yshift=-5pt},
      xlabel style = {yshift=2pt},
      grid=both,
      ymode=log,
      legend style={
  at={(0.67,1),
  font=\fontsize{5}{5.5}\selectfont},
  anchor=north west,
  draw=none,          
  fill=none,          
  cells={align=left},  
  }
  ]
    
    \addplot [red,dashed,line width=1pt] table [x index=0,
    y expr=\thisrowno{1} - \minybrainOne] {./figs/MRI/Radial/Knee1/REDEnergy_QNP/lst_cost_time.txt};
    
    \addplot [blue,dash dot,line width=1pt] table [x index=0,y expr=\thisrowno{1}-\minybrainOne] {./figs/MRI/Radial/Knee1/REDEnergy_FISTA_LL/lst_cost_time.txt};

    \addplot [green,solid,line width=1pt] table [y expr=\thisrowno{1}-\minybrainOne, x index=0] {./figs/MRI/Radial/Knee1/REDEnergy_ISTA/lst_cost_time.txt};
  
     \addplot [black,dotted,line width=1pt] table [y expr=\thisrowno{1}-\minybrainOne, x index=0] {./figs/MRI/Radial/Knee1/REDEnergy_GD/lst_cost_time.txt};

  \end{axis}
  \node[anchor=north west, xshift= 42pt, yshift=-18pt] at (CostTime.south west) {(b)};
  
\begin{axis}[
      name=PSNRIter,
     at={(CostIter.south)},
    anchor=north,
	width=0.28\textwidth,
	yshift = -36pt,
 xtick={0, 30, 60, 90,120},
  xticklabels={0, 30, 60, 90,120},
  ytick={20, 28, 35, 45},
 yticklabels={20, 28, 35, 45},
  	xmin=0, xmax=120,
  	ymax=45,
    ymin=19, 
      xlabel={Iteration},
      ylabel={PSNR (dB)},
      tick label style={font=\fontsize{5}{5.5}\selectfont},
      label style={font=\fontsize{7}{8.4}\selectfont},
      grid=both,
      legend style={
  at={(0.67,0.36),
  font=\fontsize{5}{5.5}\selectfont},
  anchor=north west,
  draw=none,          
  fill=none,          
  cells={align=left},  
  }
  ]
    
    \addplot [red,dashed,line width=1pt] table [y index=1, x expr=\coordindex] {./figs/MRI/Radial/Knee1/REDEnergy_QNP/lst_psnr_time.txt};

    \addplot [blue,dash dot,line width=1pt] table [y index=1, x expr=\coordindex] {./figs/MRI/Radial/Knee1/REDEnergy_FISTA_LL/lst_psnr_time.txt};

    \addplot [green,solid,line width=1pt] table [y index=1, x expr=\coordindex] {./figs/MRI/Radial/Knee1/REDEnergy_ISTA/lst_psnr_time.txt};

  \addplot [black,dotted,line width=1pt] table [y index=1, x expr=\coordindex] {./figs/MRI/Radial/Knee1/REDEnergy_GD/lst_psnr_time.txt};

  \end{axis}
  \node[anchor=north west, xshift= 42pt, yshift=-18pt] at (PSNRIter.south west) {(c)};

  \begin{axis}[
      name=PSNRTime,
     at={(CostTime.south)},
    anchor=north,
    yshift = -36pt,
	width=0.28\textwidth,
  yticklabels={},
   xtick={0, 15, 30, 45},
  xticklabels={0, 15, 30, 45},
  	xmin=0, xmax=45,
    ymin=19, ymax=45,
     xlabel={Wall Time (Seconds)},
      tick label style={font=\fontsize{5}{5.5}\selectfont},
      label style={font=\fontsize{7}{8.4}\selectfont},
      grid=both,
      legend style={
  at={(0.62,0.42),
  font=\fontsize{7}{8.4}\selectfont},
  anchor=north west,
  draw=none,          
  fill=none,          
  cells={align=left},  
  }
  ]
    
    \addplot [red,dashed,line width=1pt] table [y index=1, x index=0] {./figs/MRI/Radial/Knee1/REDEnergy_QNP/lst_psnr_time.txt};
        
    \addplot [blue,dash dot,line width=1pt] table [y index=1, x index=0] {./figs/MRI/Radial/Knee1/REDEnergy_FISTA_LL/lst_psnr_time.txt};
    
    \addplot [green,solid,line width=1pt] table [y index=1, x index=0] {./figs/MRI/Radial/Knee1/REDEnergy_ISTA/lst_psnr_time.txt};
    
     \addplot [black,dotted,line width=1pt] table [y index=1, x index=0] {./figs/MRI/Radial/Knee1/REDEnergy_GD/lst_psnr_time.txt};
  
  \end{axis}
  \node[anchor=north west, xshift= 42pt, yshift=-18pt] at (PSNRTime.south west) {(d)};
\end{tikzpicture}
\caption{Comparison of different methods with radial acquisition on the knee image
for $\varepsilon = 8\times10^{-5}$.
(a), (b): cost values versus iteration and wall time;
(c), (d): PSNR values versus iteration and wall time.}
\label{fig:RadialKnee1CostPSNR}
\end{figure}
\input{figs/RadialKnee1RecoIm}
\Cref{fig:RadialKnee1CostPSNR} summarizes the cost and PSNR values of each approach with respect to the number of iterations and wall time for the reconstruction with radial acquisition on the knee image. It is apparent that the performance trends here are consistent with those observed in the reconstruction with spiral acquisition on the brain image. \Cref{fig:RadialKnee1:visual} presents the reconstructed images and error maps of each method at the $50$ and $100$th iterations. It is evident that our method outperformed other methods.
The supplementary material summarizes
the results of other additional test images,
where similar trends were observed.

\subsection{Cartesian Acquisition Reconstruction}
\label{sec:numericalExp:Cartesian}

\begin{figure}[t]
	\centering
	\begin{tikzpicture}
  \pgfmathsetmacro{\minybrainOne}{-0.68+7e-5} 
  \begin{axis}[
      name=CostIter,
     at={(0,0)},
    anchor=south west,
	width=0.28\textwidth,
 xtick={0, 100,200,300,350},
  xticklabels={0,100,200,300,350},
 	xmin=0, xmax=350,
      xlabel={Iteration},
      ylabel={\MRcb{$F(\uvx_k)-F^*+\varepsilon$}},
        tick label style={font=\fontsize{5}{5.5}\selectfont},
      label style={font=\fontsize{7}{8.4}\selectfont},
      ylabel style = {yshift=-5pt},
      xlabel style = {yshift=2pt},
      grid=both,
      ymode=log,
      legend style={
  at={(0.5,1),
  font=\fontsize{7}{8.4}\selectfont},
  anchor=north west,
  draw=none,          
  fill=none,          
  cells={align=left},  
  }
  ]

     \addplot [black,dotted,line width=1pt] table [y expr=\thisrowno{1}-\minybrainOne, x expr=\coordindex] {./figs/MRI/Cartesian/DeepBrain1/NewRun/REDEnergy_GD/lst_timecost.txt};
    \addlegendentry{GD}

    \addplot [green,solid,line width=1pt] table [y expr=\thisrowno{1}-\minybrainOne, x expr=\coordindex] {./figs/MRI/Cartesian/DeepBrain1/NewRun/REDEnergy_ISTA/lst_timecost.txt};
     \addlegendentry{PG}
     
       \addplot [blue,dash dot,line width=1pt] table [x expr=\coordindex,y expr=\thisrowno{1}-\minybrainOne] {./figs/MRI/Cartesian/DeepBrain1/NewRun/REDEnergy_FISTA_LL/lst_timecost.txt};
     \addlegendentry{APG}
     
    \addplot [red,dashed,line width=1pt] table [x expr=\coordindex, y expr=\thisrowno{1}-\minybrainOne] {./figs/MRI/Cartesian/DeepBrain1/NewRun/REDEnergy_QNP/lst_timecost.txt};
    \addlegendentry{CQNPM}

%
%
%
%

  \end{axis}
  \node[anchor=north west, xshift= 42pt, yshift=-18pt] at (CostIter.south west) {(a)};

   \begin{axis}[
      name=CostTime,
        at={(CostIter.south east)},
    anchor=south west,
	width=0.28\textwidth,
	xshift=10pt,
 xtick={0, 4, 8, 12},
  xticklabels={0, 4, 8, 12},
yticklabels={},
 	xmin=0, xmax=12,
      xlabel={Wall Time (Seconds)},
      tick label style={font=\fontsize{5}{5.5}\selectfont},
      label style={font=\fontsize{7}{8.4}\selectfont},
      ylabel style = {yshift=-5pt},
      xlabel style = {yshift=2pt},
      grid=both,
      ymode=log,
      legend style={
  at={(0.67,1),
  font=\fontsize{7}{8.4}\selectfont},
  anchor=north west,
  draw=none,          
  fill=none,          
  cells={align=left},  
  }
  ]
    
   \addplot [red,dashed,line width=1pt] table [x index=0,
    y expr=\thisrowno{1} - \minybrainOne] {./figs/MRI/Cartesian/DeepBrain1/NewRun/REDEnergy_QNP/lst_timecost.txt};
    
    \addplot [blue,dash dot,line width=1pt] table [x index=0,y expr=\thisrowno{1}-\minybrainOne] {./figs/MRI/Cartesian/DeepBrain1/NewRun/REDEnergy_FISTA_LL/lst_timecost.txt};

    \addplot [green,solid,line width=1pt] table [y expr=\thisrowno{1}-\minybrainOne, x index=0] {./figs/MRI/Cartesian/DeepBrain1/NewRun/REDEnergy_ISTA/lst_timecost.txt};
  
     \addplot [black,dotted,line width=1pt] table [y expr=\thisrowno{1}-\minybrainOne, x index=0] {./figs/MRI/Cartesian/DeepBrain1/NewRun/REDEnergy_GD/lst_timecost.txt};
     
%
%
%
%

  \end{axis}
  \node[anchor=north west, xshift= 42pt, yshift=-18pt] at (CostTime.south west) {(b)};
  
\begin{axis}[
      name=PSNRIter,
     at={(CostIter.south)},
    anchor=north,
	width=0.28\textwidth,
	yshift = -36pt,
      tick label style={font=\fontsize{5}{5.5}\selectfont},
      label style={font=\fontsize{7}{8.4}\selectfont},
 xtick={0, 100,200,300,350},
  xticklabels={0, 100,200,300,350},
   ytick={21, 29, 38.5},
  yticklabels={21, 29, 38.5},
  	xmin=0, xmax=350,
    ymin=21, ymax=38.5,
      xlabel={Iteration},
      ylabel={PSNR (dB)},
      tick label style={font=\fontsize{5}{5.5}\selectfont},
      label style={font=\fontsize{7}{8.4}\selectfont},
      grid=both,
      legend style={
  at={(0.67,0.36),
  font=\fontsize{5}{5.5}\selectfont},
  anchor=north west,
  draw=none,          
  fill=none,          
  cells={align=left},  
  } 
  ]

  \addplot [red,dashed,line width=1pt] table [y index=1, x expr=\coordindex] {./figs/MRI/Cartesian/DeepBrain1/NewRun/REDEnergy_QNP/lst_timepsnr.txt};
 
    \addplot [blue,dash dot,line width=1pt] table [y index=1, x expr=\coordindex] {./figs/MRI/Cartesian/DeepBrain1/NewRun/REDEnergy_FISTA_LL/lst_timepsnr.txt};
   
    \addplot [green,solid,line width=1pt] table [y index=1, x expr=\coordindex] {./figs/MRI/Cartesian/DeepBrain1/NewRun/REDEnergy_ISTA/lst_timepsnr.txt};
  
     \addplot [black,dotted,line width=1pt] table [y index=1, x expr=\coordindex] {./figs/MRI/Cartesian/DeepBrain1/NewRun/REDEnergy_GD/lst_timepsnr.txt};
     
%
%
%
%

  \end{axis}
  \node[anchor=north west, xshift= 42pt, yshift=-18pt] at (PSNRIter.south west) {(c)};

  \begin{axis}[
      name=PSNRTime,
     at={(CostTime.south)},
    anchor=north,
    yshift = -36pt,
	width=0.28\textwidth,
  xtick={0, 4, 8,12},
  xticklabels={0, 4, 8,12},
  yticklabels={},
  	xmin=0, xmax=12,
      ymin=21, ymax=38.5,
     xlabel={Wall Time (Seconds)},
      tick label style={font=\fontsize{5}{5.5}\selectfont},
      label style={font=\fontsize{7}{8.4}\selectfont},
      grid=both,
      legend style={
  at={(0.67,0.36),
  font=\fontsize{5}{5.5}\selectfont},
  anchor=north west,
  draw=none,          
  fill=none,          
  cells={align=left},  
  }
  ]

      \addplot [red,dashed,line width=1pt] table [y index=1, x index=0] {./figs/MRI/Cartesian/DeepBrain1/NewRun/REDEnergy_QNP/lst_timepsnr.txt};
 
    \addplot [blue,dash dot,line width=1pt] table [y index=1, x index=0] {./figs/MRI/Cartesian/DeepBrain1/NewRun/REDEnergy_FISTA_LL/lst_timepsnr.txt};
    
    \addplot [green,solid,line width=1pt] table [y index=1, x index=0] {./figs/MRI/Cartesian/DeepBrain1/NewRun/REDEnergy_ISTA/lst_timepsnr.txt};
     
     \addplot [black,dotted,line width=1pt] table [y index=1, x index=0] {./figs/MRI/Cartesian/DeepBrain1/NewRun/REDEnergy_GD/lst_timepsnr.txt};

%
%
%

  \end{axis}
  \node[anchor=north west, xshift= 42pt, yshift=-18pt] at (PSNRTime.south west) {(d)};
\end{tikzpicture}
\caption{\MRcb{Comparison of different methods with Cartesian acquisition on the brain image
for $\varepsilon = 9 \times 10^{-4}$.
(a), (b): cost values versus iteration and wall time;
(c), (d): PSNR values versus iteration and wall time.}}
\label{fig:CartesianBrain1CostPSNR}
\end{figure}
\input{figs/CartesianBrain1RecoIm}

\Cref{fig:CartesianBrain1CostPSNR} shows the cost and PSNR values of each method with respect to the number of iterations and wall time for the reconstruction with Cartesian acquisition on the brain image. It is evident that similar trends were identified. Moreover, we observed that the difference between GD and PG in terms of wall time was smaller than in the cases of spiral and radial acquisitions. This is because computing $\umA\uvx$ is less expensive in this case than in the non-Cartesian acquisitions, resulting in a more efficient evaluation of the proximal mapping. \Cref{fig:CartesianBrain1:visual} shows the reconstructed images and error maps at the $50$ and $100$th iterations. There is no doubt that our method yielded a clearer image than the others.
The supplementary material shows
additional experimental results of five other brain images,
where similar trends were obtained.

\subsection{Convergence Validation}
\label{sec:numericalExp:ConvVal}

We studied the convergence properties of our method experimentally.
Let $E(\uvx_k) = \|\uvx_k - \uvx_{k+1}\|_2^2$.
\Cref{fig:ConvVal} shows the cost values and the normalized difference $E(\uvx_k)/E(\uvx_1)$ of our method with spiral acquisition on all brain test images.
As expected, the cost values converged to a constant for all test images,
and $E(\uvx_k)/E(\uvx_1) \rightarrow 0$,
consistent with our theoretical analysis.
\Cref{fig:ConvVal:Exp} presents the expected cost values versus iteration,
estimated using the Monte Carlo method with $1000$ samples.
We observed that the expected cost values decreased
similarly to \Cref{fig:ConvVal}~(a), though slightly more slowly,
aligning well with our theoretical results.

\begin{figure}[t]
    \centering
\begin{tikzpicture}
\pgfplotsset{set layers}
  \begin{axis}[
      name=CostIter,
     at={(0,0)},
     xshift=-20pt,
    anchor=south west,
	width=0.27\textwidth,
 	xmin=0, xmax=300,
      xlabel={Iteration},
      ylabel={$F(\uvx_k)/F(\uvx_1)$},
      tick label style={font=\fontsize{5}{5.5}\selectfont},
      label style={font=\fontsize{7}{8.4}\selectfont},
      ylabel style = {yshift=-5pt},
      xlabel style = {yshift=2pt},
      grid=both,
      legend style={
  at={(0.6,1),
  font=\fontsize{5}{5.5}\selectfont},
  anchor=north west,
  draw=none,          
  fill=none,          
  cells={align=left},  
  }
  ]
  
\addplot[name path=lower,dashed, red,draw=none] table [x expr=\coordindex, y expr=\thisrowno{0}] {./figs/SpiralBrain_cost_min.txt};
  
\addplot[name path=upper,dashed,blue,draw=none] table [x expr=\coordindex, y expr=\thisrowno{0}] {./figs/SpiralBrain_cost_max.txt};  

\tikzfillbetween[of=lower and upper,on layer=axis background]{fill=red!20};

\addplot [red,dashed,line width=1pt] table [x expr=\coordindex, y expr=\thisrowno{0}] {./figs/SpiralBrain_cost_mean.txt};
\end{axis}
\node[anchor=north west, xshift= 20pt, yshift=-18pt] at (CostIter.south west) {(a)};
  
  \begin{axis}[
      name=IterNormIter,
        at={(CostIter.south east)},
    anchor=south west,
    ylabel = {$E(\uvx_k)/E(\uvx_1)$},
	width=0.27\textwidth,
	xshift=15pt,
 	xmin=0, xmax=300,
      xlabel={Iteration},
      tick label style={font=\fontsize{5}{5.5}\selectfont},
      label style={font=\fontsize{7}{8.4}\selectfont},
      ylabel style = {yshift=-5pt},
      xlabel style = {yshift=2pt},
      grid=both,
      ymode=log,
      legend style={
  at={(0.67,1),
  font=\fontsize{5}{5.5}\selectfont},
  anchor=north west,
  draw=none,          
  fill=none,          
  cells={align=left},  
  }
  ]
    
 \addplot[name path=lower,dashed, red,draw=none] table [x expr=\coordindex, y expr=\thisrowno{0}] {./figs/SpiralBrain_iter_norm_min.txt};
  
\addplot[name path=upper,dashed,blue,draw=none] table [x expr=\coordindex, y expr=\thisrowno{0}] {./figs/SpiralBrain_iter_norm_max.txt};  

\tikzfillbetween[of=lower and upper,on layer=axis background]{fill=red!20};

\addplot [red,dashed,line width=1pt] table [x expr=\coordindex, y expr=\thisrowno{0}] {./figs/SpiralBrain_iter_norm_mean.txt};
  \end{axis}
\node[anchor=north west, xshift= 55pt, yshift=-18pt] at (IterNormIter.south west) {(b)};
  
  \end{tikzpicture}
    \caption{Averaged cost values (a) and $E(\uvx_k)/E(\uvx_1)$ (b) versus iteration for the proposed method. The shaded region of each curve represents the range of the cost values and $E(\uvx_k)$
    across six brain test images with spiral acquisition.}
    \label{fig:ConvVal}
\end{figure}

\begin{figure}[t]
    \centering
   \begin{tikzpicture}
\pgfplotsset{set layers}
  \begin{axis}[
      name=CostIter,
     at={(0,0)},
    anchor=south west,
	width=0.43\textwidth,
 	xmin=0, xmax=300,
      xlabel={Iteration},
      ylabel={$\mathbb{E}\left[F(\uvx_k)\right]/F(\uvx_1)$},
      tick label style={font=\fontsize{5}{5.5}\selectfont},
      label style={font=\fontsize{8}{9.2}\selectfont},
      ylabel style = {yshift=-5pt},
      xlabel style = {yshift=2pt},
      grid=both,
      legend style={
  at={(0.6,1),
  font=\fontsize{5}{5.5}\selectfont},
  anchor=north west,
  draw=none,          
  fill=none,          
  cells={align=left},  
  }
  ]
  
\addplot[name path=lower,dashed, red,draw=none] table [x expr=\coordindex, y expr=\thisrowno{0}] {./figs/UniformSpiralBrain_cost_min.txt};
  
\addplot[name path=upper,dashed,blue,draw=none] table [x expr=\coordindex, y expr=\thisrowno{0}] {./figs/UniformSpiralBrain_cost_max.txt};  

\tikzfillbetween[of=lower and upper,on layer=axis background]{fill=red!20};

\addplot [red,dashed,line width=1pt] table [x expr=\coordindex, y expr=\thisrowno{0}] {./figs/UniformSpiralBrain_cost_mean.txt};
\end{axis}
  \end{tikzpicture}
    \caption{Averaged expected cost values versus iteration for the proposed method. The shaded region of each curve represents the range of expected cost values across six brain test images with spiral acquisition. The cost values were estimated using the Monte Carlo method with $1000$ samples.}
    \label{fig:ConvVal:Exp}
\end{figure}


\subsection{\MRcb{Comparison with the Preconditioned PnP Approach}}
\label{sec:numericalExp:CompLBFGSandISTA}
\MRcb{We studied the difference between using gradient-driven denoisers
and classical CNN-based denoisers within the PnP framework.
Specifically, we adopted the preconditioned PnP method
($\text{P}^2$nP)~\cite{hong2024provable},
as it generally outperforms the classical PnP-ISTA method.
\Cref{fig:SpiralRadialKnee1PSNR} reports the PSNR values versus iterations
for CQNPM and $\text{P}^2$nP under spiral (respectively, radial) acquisition
for the brain (respectively, knee) image.
We observed that CQNPM converged faster than $\text{P}^2$nP
and consistently achieved higher PSNR values,
illustrating that the use of gradient-driven denoisers
does not sacrifice reconstruction performance.

\begin{figure}[t]
	\centering
	\begin{tikzpicture}
  \pgfmathsetmacro{\minybrainOne}{-13.73+0.3e-4} 
\begin{axis}[
      name=PSNRIterSpiralBrain,
    at={(0,0)},
    anchor=north,
	width=0.28\textwidth,
	yshift = -36pt,
 xtick={0, 50, 100, 150},
  xticklabels={0, 50, 100, 150},
   ytick={20, 28, 35, 41},
  yticklabels={20, 28, 35, 41},
  	xmin=0, xmax=150,
    ymin=19, ymax=41,
      xlabel={Iteration},
      ylabel={PSNR (dB)},
       tick label style={font=\fontsize{5}{5.5}\selectfont},
      label style={font=\fontsize{7}{8.4}\selectfont},
      grid=both,
      legend style={
  at={(0.42,0.4),
  font=\fontsize{6}{6.5}\selectfont},
  anchor=north west,
  draw=none,          
  fill=none,          
  cells={align=left},  
  }
  ]
    
    \addplot [red,dashed,line width=1pt] table [y index=1, x expr=\coordindex] {./figs/MRI/Spiral/DeepBrain1/REDEnergy_QNP/lst_psnr_time.txt};
   \addlegendentry{CQNPM}
    
    \addplot [blue,dash dot,line width=1pt] table [y index=1, x expr=\coordindex] {./figs/MRI/lst_psnr_time_PPnPSpiralBrain1.txt};
   			\addlegendentry{$\text{P}^2$nP}
  \end{axis}
  \node[anchor=north west, xshift= 42pt, yshift=-18pt] at (PSNRIterSpiralBrain.south west) {(a)};

\begin{axis}[
      name=PSNRIter,
     at={(PSNRIterSpiralBrain.south east)},
     anchor=south west,
	width=0.28\textwidth,
	xshift=10pt,
 xtick={0, 40, 80, 120},
  xticklabels={0, 40, 80, 120},
   ytick={20, 28, 35, 45},
  yticklabels={20, 28, 35, 45},
  	xmin=0, xmax=120,
    ymin=19, ymax=45,
      xlabel={Iteration},
       tick label style={font=\fontsize{5}{5.5}\selectfont},
      label style={font=\fontsize{7}{8.4}\selectfont},
      grid=both,
      legend style={
  at={(0.67,0.36),
  font=\fontsize{5}{5.5}\selectfont},
  anchor=north west,
  draw=none,          
  fill=none,          
  cells={align=left},  
  }
  ]
    
    \addplot [red,dashed,line width=1pt] table [y index=1, x expr=\coordindex] {./figs/MRI/Radial/Knee1/REDEnergy_QNP/lst_psnr_time.txt};

    \addplot [blue,dash dot,line width=1pt] table [y index=1, x expr=\coordindex] {./figs/MRI/lst_psnr_time_PPnPRadialKnee1.txt};

  \end{axis}
  \node[anchor=north west, xshift= 42pt, yshift=-18pt] at (PSNRIter.south west) {(b)};

  \end{tikzpicture}
\caption{Comparison of the preconditioned PnP (($\mathrm{P}^2$PnP)) method.  
(a) PSNR values versus iteration for spiral acquisition with the brain image;  
(b) PSNR values versus iteration for radial acquisition with the knee image.}
\label{fig:SpiralRadialKnee1PSNR}
\end{figure}

}

\section{\MRcb{Conclusion and Future Work}}
\label{sec:conclusion}

Compared to the PnP/RED framework with convolutional neural network based denoisers,
the gradient-driven denoisers offer a much stronger theoretical foundation,
as its required assumption (i.e., Lipschitz continuity of $f(\uvx)$)
is easier to satisfy in practice—--%
an important advantage for medical imaging applications.
We applied the gradient-driven denoisers to CS MRI reconstruction
and thoroughly evaluated its performance on spiral, radial, and Cartesian acquisitions.
In addition, we proposed a \emph{complex} quasi-Newton proximal method (CQNPM) to efficiently solve the associated minimization problem 
with convergence guarantee under nonconvex settings.
We extensively compared our method with existing algorithms,
and the experimental results demonstrate both the efficiency of our approach
and the accuracy of the underlying theoretical analysis.
Although we consider $h(\uvx)=\frac{1}{2}\|\umA\uvx-\uvy\|_2^2$ in this paper,
the algorithm presented is applicable to any $h$
that is convex and smooth (Lipschitz gradient).
Furthermore,
our theoretical results require only the convexity of $h$,
so the approach also generalizes to non-smooth and convex $h$
by modifying the weighted proximal mapping 
at step~\ref{alg:PropMethod:WPG} in \Cref{alg:PropMethod} appropriately.
\MRcb{Code to reproduce the results in the paper
is available at
\url{https://github.com/hongtao-argmin/CQNPM-GDD-CS-MRI-Reco}.}


\MRcb{
While the proposed CQNPM accelerates convergence
in gradient-driven denoisers based CS MRI reconstruction, several limitations remain.
In the following, we outline these limitations and discuss possible directions for improvement:

\begin{itemize}
\item \textbf{Accurate Hermitian positive definite Hessian Approximation:}
In this work, we employed~\Cref{alg:ModfiedZero:SR1}
to estimate a Hermitian positive definite Hessian matrix,
but we always reinitialized the Hessian at each step.
In the real-valued setting,
BFGS updating typically provides more accurate Hessian approximations
compared to the scheme in~\Cref{alg:ModfiedZero:SR1}.
Therefore, extending BFGS to our problem—while ensuring that the Hessian matrix remains Hermitian positive definite when $f$ is nonconvex—could potentially yield a faster algorithm.

\item \textbf{Assumptions for convergence:}  
The convergence rate of our method depends on the proximal PL inequality.
However, verifying this condition for complex, high-capacity neural network–based denoisers
remains challenging in practice.
Extending the analysis to the more general KL inequality,
or alternatively designing convex networks~\cite{amos2017input}
that still achieve comparable performance, would be promising directions for future research.

\item \textbf{Extension to 3D and dynamic MRI reconstruction:}  
This work focuses on 2D multi-coil CS MRI reconstruction.
Extending CQNPM to 3D or dynamic MRI reconstruction would be an interesting future direction.
However, training effective gradient-driven denoisers for such high-dimensional problems
remains a significant challenge that must be addressed.


\end{itemize}

}
\appendices
\crefalias{section}{appendix}


\section{Proof of \texorpdfstring{\Cref{lemma:D_h:GM:inequality}}{Lemma 2}}
\label{proof:lemma:D_h:GM:inequality}

Since $\umW\succ 0$ and $h(\uvx)$ is convex,
$\mathcal S_{h}(\bar{\uvx},\uvx,\uvg,\umW,\stepsize)$ is strongly convex.
Through the optimality condition of \eqref{eq:OptimalSoluS_h},
we have the following inequality
$$
\Re\left\{\left\langle \uvg+\frac{1}{\stepsize}\umW(\bar{\uvx}^+-\uvx)+\nabla h(\bar{\uvx}),\uvx-\bar{\uvx}^+\right\rangle\right\}\geq 0,~~\forall  \uvx\in\mathcal C.
$$
By using the convexity of $h$ (i.e., $h(\uvx)-h(\bar{\uvx})\geq \Re\{ \langle \nabla h(\bar{\uvx}),\uvx-\bar{\uvx}\rangle\}$), we reach
\begin{equation}
\label{eq:FirstOrderOptSh:Inequality}
\Re\left\{\left\langle \uvg,\uvx-\bar{\uvx}^+\right\rangle\right\}\geq \frac{1}{\stepsize} \|\bar{\uvx}^+-\uvx\|_{\umW}^2+h(\uvx^+)-h(\uvx).    
\end{equation}

From \eqref{eq:def:D_h} and \eqref{eq:OptimalSoluS_h}, we have
$$
\begin{array}{rcl}
\mathcal D_{h}^{\mathcal C}(\uvx,\uvg,\umW,\stepsize)&=&
-\frac{2}{\stepsize}\mathcal S_{h}(\bar{\uvx}^+,\uvx,\uvg,\umW,\stepsize)\\[5pt]
&=&\frac{2}{\stepsize}\Re\big\{\langle \uvg,\uvx-\bar{\uvx}^+ \rangle\big\} -\frac{1}{\stepsize^2}\|\bar{\uvx}^+-\uvx\|_{\umW}^2\\[3pt]
&&\quad\quad\quad-\frac{2}{\stepsize}(h(\bar{\uvx}^+)-h(\uvx))\\
&\geq & \frac{1}{\stepsize^2}\|\bar{\uvx}^+-\uvx\|_{\umW}^2\\
&=&\|\mathcal G_{\frac{1}{\stepsize},\umW}^{f,h}(\uvx)\|_{\umW}^2,
\end{array}
$$
where the first inequality and last equality  come from \eqref{eq:FirstOrderOptSh:Inequality} and the definition of $\mathcal G_{\frac{1}{\stepsize},\umW}^{f,h}(\uvx)$.

\section{Proof of \texorpdfstring{\Cref{lemma:Property:D_h:stepsize}}{Lemma 3}}
\label{proof:lemma:Property:D_h:stepsize}

Let $\stepsize_1\geq \stepsize_2>0$ and $\bar{\uvx}_2=\frac{\stepsize_2}{\stepsize_1}\bar{\uvx}_1+\frac{\stepsize_1-\stepsize_2}{\stepsize_1}\uvx$ with $\bar{\uvx}_1,\bar{\uvx}_2\in\mathcal C$. Since $h(\uvx)$ is convex, we get
$$
h(\bar{\uvx}_2)\leq \frac{\stepsize_2}{\stepsize_1}h(\bar{\uvx}_1)+\frac{\stepsize_1-\stepsize_2}{\stepsize_1}h(\uvx)
$$
and then
$$
\stepsize_1(h(\bar{\uvx}_2)-h(\uvx))\leq \stepsize_2(h(\bar{\uvx}_1)-h(\uvx)).
$$
By using $\stepsize_1(\bar{\uvx}_2-\uvx)=\stepsize_2(\bar{\uvx}_1-\uvx)$ and the above inequality, we have
$$
-\frac{2}{\stepsize_1}\mathcal S_{h}(\bar{\uvx}_1,\uvx,\uvg,\umW,\stepsize_1)\leq -\frac{2}{\stepsize_2}\mathcal S_{h}(\bar{\uvx}_2,\uvx,\uvg,\umW,\stepsize_2).
$$
Minimizing both sides of the above inequalities and using  the definition of $\mathcal D_{h}^{\mathcal C}(\uvx,\uvg,\umW,\stepsize)$, we get the desired result
$$ \mathcal D_{h}^{\mathcal C}(\uvx,\uvg,\umW,\stepsize_1)\leq \mathcal D_{h}^{\mathcal C}(\uvx,\uvg,\umW,\stepsize_2).$$ 

\section{Proof of \texorpdfstring{\Cref{lemma:boundedHessian}}{Lemma 4}}
\label{proof:lemma:boundedHessian}

To prove the bound on $\umB_k$,
we first establish the bound on $\umH_k$ because $\umB_k=\umH_k^{-1}$.
Define $a_k=\langle \uvs_k,\uvs_k \rangle, \
b_k=\Re\{\langle \uvs_k,\uvv_k \rangle\}, \
c_k=\langle \uvv_k,\uvv_k \rangle.$
Then we have
$$
\frac{b_k}{a_k}\tau_k = 1-\sqrt{1-\frac{b_k^2}{a_k\,c_k}}\leq \frac{b_k^2}{a_k\,c_k},
$$
resulting in $\tau_k\leq \frac{b_k}{c_k}$.
The last inequality follows from the Cauchy–Schwarz inequality,
which leads to $b_k^2\leq a_k c_k$.
The lower bound can be derived through
\begin{align*}
\tau_k=\frac{a_k}{b_k}-\sqrt{\Big(\frac{a_k}{b_k}\Big)^2-\frac{a_k}{c_k}}
=&
\frac{\frac{a_k}{c_k}}{\frac{a_k}{b_k}+\sqrt{\Big(\frac{a_k}{b_k}\Big)^2-\frac{a_k}{c_k}}}
\\
>&
\frac{b_k}{2c_k}.
\end{align*}
In summary, we have
$\frac{b_k}{2c_k}\leq \tau_k \leq \frac{b_k}{c_k}$.
Next, we derive an upper bound for $\uvu_k^\HTrans\uvu_k$.
Using the definition of $\uvu_k$, we have the following inequalities
$$
\begin{array}{rcl}
\uvu_k^\HTrans\uvu_k
&=&\frac{\langle \uvs_k-\tau_k\uvv_k,\uvs_k-\tau_k\uvv_k \rangle}{\rho_k}\\[3pt]
&\leq & \frac{\|\uvs_k-\tau_k\uvv_k\|_2}{\delta \|\uvv_k\|_2}\\
&=&\frac{1}{\delta}\sqrt{\frac{a_k}{c_k}-\frac{2\tau_k b_k}{c_k}+\tau_k^2}\\
&\leq& \frac{1}{\delta}\sqrt{\frac{a_k}{c_k}}=\frac{1}{\delta}\sqrt{\frac{a_k}{b_k} \frac{b_k}{c_k}}\leq \frac{a_k}{\delta b_k}. 
\end{array}
$$
The first inequality comes from step
\ref{alg:ModfiedZero:SR1:u_k0} in \Cref{alg:ModfiedZero:SR1}.
The second inequality derived from the fact that
$\frac{b_k}{2c_k}\leq \tau_k\leq \frac{b_k}{c_k}$
resulting in $-\frac{2\tau_k b_k}{c_k}+\tau_k^2<0$.
The last inequality is the result of $b_k^2\leq a_k c_k$. 

With these, we have 
$
\frac{b_k}{2c_k} \umI_N\preceq \umH_k\preceq (\frac{b_k}{c_k}+\frac{a_k}{\delta b_k})\umI_N.
$
From \eqref{eq:alg:Zero:SC:SR1:alpha}, we know $\frac{b_k}{a_k}\geq \theta_1$ and $\frac{c_k}{b_k}\leq \theta_2$ for all $k$. Therefore, $\umH_k$ is always bounded by 
$
\frac{1}{2\theta_2} \umI_N\preceq \umH_k\preceq (\frac{1+\delta}{\delta \theta_1})\umI_N.
$
Since $\umH_k$ is both lower and upper bounded, we know that there exist constants $\overline{\eta} > \underline{\eta} > 0$ such that
$
\underline{\eta}\,\umI \preceq \umB_k \preceq \overline{\eta}\,\umI.
$


\section{Proof of \texorpdfstring{\Cref{them:ConvResults}}{Thm 1}}
\label{proof:them:ConvResults}

Using \Cref{lemma:DescentLemma}, we have 
\begin{align*}
f(\uvx_{k+1}) &\leq
f(\uvx_k)+\Re\big\{\langle \nabla f(\uvx_{k}),\uvx_{k+1}-\uvx_k\rangle\big\}
\\&
\quad+\frac{L}{2}\|\uvx_{k+1}-\uvx_k\|_2^2
\\ &\leq
f(\uvx_k)+h(\uvx_k)-h(\uvx_{k+1})
\\ & \quad
+ \min\limits_{\uvv\in\mathcal C}
\mathcal S_h(\uvv,\uvx_k,\nabla f(\uvx_k),\umB_k,\stepsize_k)
\\ & \quad
+ \frac{L}{2}\|\uvx_{k+1}-\uvx_k\|_2^2-\frac{1}{2\stepsize_k}
\|\uvx_{k+1}-\uvx_k\|_{\umB_k}^2
\\ & \leq
F(\uvx_k) + \left( \min\limits_{\uvv\in\mathcal C}
\mathcal S_h(\uvv,\uvx_k,\nabla f(\uvx_k),\umB_k,\stepsize_k) \right)
\\ & \quad
+ \left( \frac{L}{2}-\frac{\underline{\eta}}{2\stepsize_k} \right)
\|\uvx_{k+1}-\uvx_k\|_2^2 -h(\uvx_{k+1}).
\end{align*}
The second inequality comes from the definition of $\mathcal S_h$
and the update rule for $\uvx_{k+1}$.
The third inequality is the result of \Cref{lemma:boundedHessian}.  
Letting $\stepsize_k\leq \frac{\underline{\eta}}{L}$
and moving $h(\uvx_{k+1})$ to the left side, we get
\begin{align}
\label{eq:proveConv:descent:F}
F(\uvx_{k+1}) &\leq
F(\uvx_k) + \left( \min\limits_{\uvv\in\mathcal C}
\mathcal S_h(\uvv,\uvx_k,\nabla f(\uvx_k),\umB_k,\stepsize_k) \right)
\nonumber \\ & \quad
+ \left( \frac{L}{2}-\frac{\underline{\eta}}{2\stepsize_k} \right)
\|\uvx_{k+1}-\uvx_k\|_2^2
\nonumber\\ & \leq
F(\uvx_k)-\frac{\stepsize_k}{2}
\mathcal D_h^{\mathcal C}(\uvx_k,\nabla f(\uvx_k),\umB_k,\stepsize_k).
\end{align}
Rearranging \eqref{eq:proveConv:descent:F}, we have
\begin{equation}
\label{eq:proveConv:ineq:D_h:H_k:stepsize_k}
\frac{\stepsize_k}{2}
\mathcal D_h^\mathcal C(\uvx_k,\nabla f(\uvx_k),\umB_k,\stepsize_k)
\leq F(\uvx_k)-F(\uvx_{k+1}).
\end{equation}
Invoking \Cref{lemma:D_h:GM:inequality,lemma:boundedHessian}, we get
\begin{equation}
\label{eq:D_h:C:gamma_k:inequality}
\mathcal D_h^\mathcal C(\uvx_k,\nabla f(\uvx_k),\umB_k,\stepsize_k)
\geq \frac{\underline{\eta}}{\stepsize_k^2}\|\uvx_{k+1}-\uvx_k\|_2^2.
\end{equation}
Substituting \eqref{eq:D_h:C:gamma_k:inequality}
into \eqref{eq:proveConv:ineq:D_h:H_k:stepsize_k}, we have
\begin{equation}
\label{eq:gamma_k:Cost:inequality}
\frac{\underline{\eta}}{2\stepsize_k}\|\uvx_{k+1}-\uvx_k\|_2^2
\leq F(\uvx_k)-F(\uvx_{k+1}).
\end{equation}
Summing up the above inequality from $k=1$ to $K$, we reach
\begin{equation}
\label{eq:sum:gamma_k:Cost:inequality}
\sum_{k=1}^K \frac{\underline{\eta}}{2\stepsize_k}
\|\uvx_{k+1}-\uvx_k\|_2^2\leq F(\uvx_1)-F(\uvx_K)\leq F(\uvx_1)-F^*.	
\end{equation}
Denote by $\stepsize_{\mathrm{max}}=\max_k \{\stepsize_k\}$, 
$\stepsize_{\mathrm{min}}=\min_k \{\stepsize_k\}$,
and $\Delta_K = \min_{k\leq K} \|\uvx_{k+1}-\uvx_k\|_2^2$.
Since $\stepsize_k\leq \frac{\underline{\eta}}{L}$,
we have $\stepsize_{\mathrm{max}}=\frac{\underline{\eta}}{L}$.
Invoking the definition of $\Delta_K$, 
the value of $\stepsize_{\mathrm{max}}$,
and using \eqref{eq:sum:gamma_k:Cost:inequality}, we obtain
$$
\Delta_K \leq \frac{2\left(F(\uvx_1)-F^*\right)}{LK}.
$$
Clearly, $\Delta_K$ approaches zero as $K\rightarrow \infty$.

Now, we show the convergence of cost values with additional \Cref{assum:PLCond}. 
Invoking \Cref{lemma:Property:D_h:stepsize,lemma:boundedHessian},
we have the following inequalities
\begin{align}
\label{eq:D_h_C:inequalities}
\mathcal D_h^\mathcal C(\uvx_k,\nabla f(\uvx_k),\umB_k,\stepsize_k)
&\geq
\frac{1}{\overline{\eta}}
\mathcal D_h^\mathcal C(\uvx_k,\nabla f(\uvx_k),\umI,\frac{\stepsize_k}{\overline{\eta}})
\nonumber\\
&\geq
\frac{1}{\overline{\eta}}
\mathcal D_h^\mathcal C(\uvx_k,\nabla f(\uvx_k),\umI,\frac{\stepsize_{\max}}{\overline{\eta}}).
\end{align}
Combining the above inequality
with \eqref{eq:proveConv:ineq:D_h:H_k:stepsize_k},
we obtain
\begin{equation}
\label{eq:proveConv:ineq:D_h:I:stepsize_k}
    \frac{\stepsize_k}{2\overline{\eta}}
    \mathcal D_h^\mathcal C(\uvx_k,\nabla f(\uvx_k),\umI,\frac{\stepsize_{\mathrm{max}}}{\overline{\eta}})
    \leq F(\uvx_k)-F(\uvx_{k+1}).
\end{equation}
By using \eqref{eq:PLIneq}, we get
$$
\frac{\nu\,\stepsize_{\mathrm{min}}}{\overline{\eta}}\,(F(\uvx_k)-F^*)
\leq F(\uvx_k)-F^*-\left(F(\uvx_{k+1})-F^*\right).
$$
Rearranging the above inequality, we reach
\begin{equation}
\label{eq:ConvFunInequalityIter}
F(\uvx_{k+1})-F^*
\leq \Big(1-\frac{\nu\,\stepsize_{\mathrm{min}}}{\overline{\eta}}\Big )
\,\big (F(\uvx_k)-F^*\big).
\end{equation}
By letting 
$\stepsize_k \leq \min\left\{\frac{\overline{\eta}}{\nu},\frac{\underline{\eta}}{L}\right\}$,
we have $\big (1-\frac{\nu\,\stepsize_{\mathrm{min}}}{\overline{\eta}}\big )>0$.
Then applying \eqref{eq:ConvFunInequalityIter} recursively, we obtain
$$
F(\uvx_{k+1})-F^*\leq \Big(1-\frac{\nu\,\stepsize_{\mathrm{min}}}{\overline{\eta}}\Big)^k\,\big (F(\uvx_1)-F^*\big).
$$

Now we show another formulation of convergence by uniformly sampling the output.
Summing up \eqref{eq:proveConv:ineq:D_h:H_k:stepsize_k} from $k=1$ to $K$, we get
\begin{equation}
\label{eq:proveConv:ineq:SumD_h:Upper}
\begin{array}{rcl}
\sum\limits_{k=1}^{K}\frac{\stepsize_k}{2}\mathcal D_h^\mathcal C(\uvx_k,\nabla f(\uvx_k),\umB_k,\stepsize_k)&\leq &F(\uvx_1)-F(\uvx_K)\\
&\leq&F(\uvx_1)-F^*
\end{array}
\end{equation}
By uniformly sampling one of the previous iterates at $K-1$th iteration
as the output $\uvx_{k'}$, we have 
\begin{equation}
\label{eq:ExpInequality}
\begin{array}{rl}
\mathbb{E}\Big[ D_h^\mathcal C(\uvx_{k'},\nabla f(\uvx_{k'}),\umI,\frac{\stepsize_{k'}}{\overline{\eta}}) \Big] =& \sum\limits_{k=1}^{K}\frac{\mathcal D_h^\mathcal C(\uvx_k,\nabla f(
    \uvx_k),\umI,\frac{\stepsize_k}{\overline{\eta}})}{K}  \\[5pt]
\leq& \sum\limits_{k=1}^{K}\frac{\overline{\eta} \mathcal D_h^\mathcal C(\uvx_k,\nabla f(
    \uvx_k),\umB_k,\stepsize_{k})}{K},
\end{array}	
\end{equation}
where the inequality comes from \eqref{eq:D_h_C:inequalities}.  Summing up \eqref{eq:proveConv:ineq:D_h:H_k:stepsize_k} from $k=1$ to $K$, we obtain
$$
\begin{array}{rcl}
	\sum_{k=1}^K \frac{\stepsize_k}{2}\mathcal D_h^\mathcal C(\uvx_k,\nabla f(\uvx_k),\umB_k,\stepsize_k)&\leq& F(\uvx_1)-F(\uvx_{K+1})\\
	&\leq& F(\uvx_1)-F^*.
\end{array}
$$
Together with \eqref{eq:ExpInequality}, we get
$$
\mathbb{E}\Big[ D_h^\mathcal C(\uvx_{k'},\nabla f(\uvx_{k'}),\umI,\frac{\stepsize_{k'}}{\overline{\eta}}) \Big]
\leq \frac{2\overline{\eta}\left(F(\uvx_1)-F^*\right)}{\stepsize_{\mathrm{min}}K}.
$$
By invoking \eqref{eq:PLIneq}, we get the desired result
$$
\mathbb{E}\Big[F(\uvx_{k'})-F^*\Big]
\leq \frac{\overline{\eta}\left(F(\uvx_1)-F^*\right)}{\nu\stepsize_{\mathrm{min}}K}. 
$$
Clearly, $F(\uvx_{k'})$ converges to $F^*$ in expectation as $K\rightarrow \infty$.

\bibliographystyle{IEEEtran}
\bibliography{Refs}

\end{document}